\documentclass[twocolumn,notitlepage,prb,superscriptaddress,longtable, longbibliography]{revtex4-2}
\usepackage[version=4]{mhchem} 
\usepackage{amsthm}
\usepackage{amsmath}
\usepackage{amssymb}
\usepackage{mathdots}
\usepackage{graphicx}
\usepackage{mathrsfs}
\usepackage{longtable}
\usepackage{dcolumn}
\usepackage{booktabs}
\usepackage{multirow}
\usepackage{physics}
\makeatletter
\@ifundefined{textcolor}{}
{%
 \definecolor{BLACK}{gray}{0}
 \definecolor{WHITE}{gray}{1}
 \definecolor{RED}{rgb}{1,0,0}
 \definecolor{GREEN}{rgb}{0,1,0}
 \definecolor{BLUE}{rgb}{0,0,1}
 \definecolor{CYAN}{cmyk}{1,0,0,0}
 \definecolor{MAGENTA}{cmyk}{0,1,0,0}
 \definecolor{YELLOW}{cmyk}{0,0,1,0}
}

\usepackage{braket}
\usepackage[backref=true,
bookmarksnumbered=true,
bookmarks=true,
bookmarksopen=true,
colorlinks=true,
citecolor=blue,
linkcolor=blue,
anchorcolor=green,
urlcolor=blue,unicode=false]{hyperref}

\begin{document}
\title{Engineering micromotion in Floquet prethermalization via space-time symmetries}
\author{Ilyoun Na}
    \email{ilyoun1214@berkeley.edu}
    \affiliation{Department of Physics, University of California, Berkeley, California 94720, USA}
    \affiliation{Materials Sciences Division, Lawrence Berkeley National Laboratory, Berkeley, California 94720, USA}
    \affiliation{Molecular Foundry, Lawrence Berkeley National Laboratory, Berkeley, California 94720, USA}
\author{Jack Kemp}
    \email{jack_kemp@g.harvard.edu}
    \affiliation{Department of Physics, Harvard University, Cambridge, Massachusetts 02138 USA}
\author{Sin\'{e}ad M. Griffin}
    \email{sgriffin@lbl.gov}
    \affiliation{Materials Sciences Division, Lawrence Berkeley National Laboratory, Berkeley, California 94720, USA}
    \affiliation{Molecular Foundry, Lawrence Berkeley National Laboratory, Berkeley, California 94720, USA}
\author{Yang Peng}
    \email{yang.peng@csun.edu}
    \affiliation{Department of Physics and Astronomy, California State University, Northridge, Northridge, California 91330, USA}
    \affiliation{Institute of Quantum Information and Matter and Department of Physics, California Institute of Technology, Pasadena, CA 91125, USA}
\date{November 25, 2025}

\begin{abstract}
We present a systematic framework for Floquet prethermalization under strong resonant driving, emphasizing the pivotal role of dynamical space-time symmetries. Our approach demonstrates how dynamical space-time symmetries map onto the projective static symmetry group of the prethermal Hamiltonian governing the prethermal regime. We introduce techniques for detecting dynamical symmetries through the time evolution of local observables, facilitating a detailed analysis of micromotion within each period and surpassing the limitations of conventional stroboscopic Floquet prethermal dynamics. To implement this framework, we present a prethermal protocol that preserves order-two dynamical symmetry in a spin-ladder model, confirming the predicted relationships between the expectation values of local observables at distinct temporal points in the Floquet cycle, linked by this symmetry. %Notably, this detection protocol imposes no constraints on the initial state, significantly enhancing experimental feasibility.
\end{abstract}

\maketitle
\section{Introduction}
Periodically driven (Floquet) systems have become a versatile platform for extending equilibrium phases to nonequilibrium settings, enabling novel phases with no equilibrium counterparts, including dynamical topological phases~\cite{Kitagawa2010,Rudner2013,Else2016_1,Titum2016,Keyserlingk2016_1,Keyserlingk2016_2,Potter2016_1,Harper2017,Po2017,Fidkowski2019} and exotic spatiotemporal phases such as discrete time crystals (DTCs)~\cite{Khemani2016,Else2016_2,Yao2017,Else2020,Zaletel2023} and quantum many-body scars~\cite{Mizuta2020,Sugiura2021,Maskara2021,Hudomal2022}. Recent experimental advancements have further facilitated the realization of these phases~\cite{Rechtsman2013,Aidelsburger2013,Miyake2013,Jotzu2014,Zhang2017,Choi2017,Wintersperger2020,Rubio-Abadal2020,Peng2021,Kyprianidis2021,Beatrez2023,Stasiuk2023}, driving active research in Floquet engineering. 

A key challenge in stabilizing these phases is managing energy absorption from periodic driving, which can lead nonintegrable many-body systems to heat up into featureless infinite-temperature states~\cite{Alessio2014,Lazarides2014}. To counteract this heating, methods that effectively suppress energy absorption through tunable external parameters are crucial. One approach employs strong disorder, inducing robust emergent integrability and confining the system to nonthermal states known as Floquet many-body localized (MBL) states~\cite{Ponte2015_1,Ponte2015_2,Lazarides2015}.

Another strategy is \textit{Floquet prethermalization}, where the system evolves into finite-temperature thermal states governed by a static effective Hamiltonian, termed the prethermal Hamiltonian, in an intermediate time regime before heating up. It has been demonstrated that Floquet many-body systems universally host long-lived prethermal quasi-equilibria in the high-frequency limit, if the local energy scale of the system remains much smaller than the driving frequency, without requiring integrability or disorder~\cite{Abanin2015,Mori2016,Kuwahara2016,Abanin2017_1,Abanin2017_2,Else2017_1,Else2017_2,Mori2018_1,Mori2018_2,Ho2023}.

However, the high-frequency limit~\cite{Bukov2015,Eckardt2015} does not encompass all intriguing Floquet phases. Notable examples, such as anomalous Floquet topological phases~\cite{Rudner2013,Titum2016,Nathan2021}, arise under moderate to strong driving, where the local energy scale can be large. By shifting to a rotating frame where the driving appears sufficiently weakened, an approximate prethermal Hamiltonian can be constructed to describe the Floquet prethermal phase~\cite{Mikami2016,Mizuta2019,Mori2022}. This approach reveals emergent symmetries in the prethermal Hamiltonian that are absent in the original Hamiltonian, paving the way for prethermal Floquet symmetry-protected topological phases~\cite{Else2017_1,Else2017_2} and prethermal DTCs~\cite{Else2020,Mizuta2019}. 

The time-dependent nature of Floquet systems permits intertwined nonsymmorphic space-time symmetries~\cite{Morimoto2017,Xu2018,Else2020_2}, combining fractional temporal translations with spatial transformations, yielding a richer topological classification than static symmetries alone in noninteracting systems~\cite{YP2019,YP2020,YP2022,Na2023}. 
Nevertheless, the effects of these dynamical symmetries are not well explored.

In this paper, we investigate space-time symmetric quantum many-body systems, focusing on their quasi-steady prethermal states, which offer a stable framework for exploring novel phases emerging from dynamical symmetries. Specifically, we identify the symmetry group $G^{int}_{st}$
of the prethermal Hamiltonian, derived from the dynamical space-time 
symmetry group $G_{st}$ preserved by the driving protocol. Building on previous studies of static symmetry groups~\cite{Else2017_1, Mizuta2019}, we rigorously extend this framework to encompass dynamical symmetries.

We further demonstrate that space-time symmetric driving protocols can be employed to engineer nontrivial micromotion during the prethermal stage. To concretize this idea, we analyze a quantum spin ladder subjected to a dual-driving protocol: one with an energy scale comparable to the driving frequency and another significantly weaker. When the driving protocol preserves an order-two dynamical symmetry, we show that the expectation values of certain observables oscillate with equal magnitude but opposite signs at integer and half-integer stroboscopic times. Crucially, this behavior is robust, independent of the initial state, and thus highly amenable to experimental realization.

\section{Dynamical space-time symmetry group}
We consider Floquet systems with Hamiltonians $\hat{H}(t)=\hat{H}(t+T)$ that exhibit a symmetry group $G_{st}$, encompassing both static and dynamical symmetries transforming both spatial and temporal coordinates~\cite{Morimoto2017,Xu2018,Else2020_2}. Within $G_{st}$, subgroups $M$ and $A$ represent unitary and antiunitary symmetries, respectively. For order-2 discrete symmetries,
under the assumption that $\hat{g}_M\in M$ and $\hat{g}_A \in A$ commute,
we have
\begin{align} \label{eq:Definition of symmetries}
    \hat{g}_M\hat{H}(t)\hat{g}^{-1}_M&=\hat{H}(t+T/2),  \\
    \hat{g}_A\hat{H}(t)\hat{g}^{-1}_A&=\hat{H}(-t+s), \nonumber
\end{align}
where $s\in\left[0,T\right]$ is a reference point within the driving period $T$. Without loss of generality, we set $s=T/2$~\cite{YP2020}. 

\section{Prethermalization in a dual-driving setup}
We consider a Hamiltonian $\hat{H}(t)=\hat{H}_0(t)+\hat{V}(t)$, where both $\hat{H}_0(t)$ and $\hat{V}(t)$ are $T$-periodic. The resonant drive $\hat{H}_0(t)$ has a local energy scale comparable to the driving frequency $\Omega=2\pi/T$, and we impose that its local terms commute at each moment of time. The weak drive $\hat{V}(t)$ has a typical energy scale $\lambda\sim 2kJ \ll \Omega$, involving interactions up to $k$-local terms. Here $J$ denotes the characteristic local interaction strength, with the cumulative interaction strength involving any site bounded by $J$~\cite{Mori2016,Kuwahara2016,SM}. The resonant drive is designed such that the one-period evolution operator generated solely by $\hat{H}_0(t)$ is 
\begin{equation} \label{eq:Definition of Symmetry X}
    \hat{U}_0(T,0)=\mathcal{T}[e^{-i\int_{0}^{T} dt\hat{H}_0(t)}]=\hat{X},\quad\hat{X}^N=\mathbb{I},
\end{equation}
which implements a spatially local $\mathbb{Z}_N$ unitary action. In the absence of $\hat{H}_0(t)$, this system lies in the high-frequency regime, while turning on $\hat{H}_0(t)$ moves it beyond this regime. 
 
To derive the prethermal static Hamiltonian for this dual-driving setup, we turn to the interaction picture with respect to $\hat{H}_0(t)$, writing $\hat{U}_0(t)\equiv\hat{U}_0(t,0)$. The interaction picture Hamiltonian is then
\begin{equation} \label{eq:Hamiltonian in the interaction picture}
    \hat{H}_{int}(t)=i\left[\partial_t\hat{U}^{\dag}_0(t)\right]\hat{U}_0(t)+\hat{U}^{\dag}_0(t)\hat{H}(t)\hat{U}_0(t).
\end{equation}
Since $\hat{H}(t)=\hat{H}_0(t)+\hat{V}(t)$, this reduces to $\hat{H}_{int}(t)=\hat{U}^{\dag}_0(t)\hat{V}(t)\hat{U}_0(t)=\hat{V}_{int}(t)$. 
From $\hat{U}_0(NT)=\mathbb{I}$ in Eq.~\eqref{eq:Definition of Symmetry X} of Symmetry X, it follows that $\hat{H}_{int}(t)$ is $NT$-periodic. 
This reformulation maps the initial strong $T$-periodic drive into a weak $NT$-periodic drive with the local energy scale $\lambda_{int}\sim\lambda$ of $\hat{V}_{int}(t)$ satisfying $1 \ll \Omega/(N\lambda)=\tilde{\Omega}$. This enables a van Vleck high-frequency expansion in terms of a unitary rotation that relates $\hat{H}_{int}(t)$ to the Floquet Hamiltonian $\hat{V}^{FM}_{int}$, which is defined through $\hat{U}_{int}(NT)=\mathcal{T}[e^{-i\int_{0}^{NT} dt\hat{V}_{int}(t)}]=e^{-i\hat{V}^{FM}_{int}NT}$. 

To perform this expansion, we introduce the rotation $\mathcal{\hat{U}}_n(t)=e^{-i\hat{K}_n(t)}$, where the kick operator $\hat{K}_n(t)$ is $NT$-periodic and expressed as $\hat{K}_n(t)=\sum^{n}_{i=1}\hat{K}^{[i]}(t)$, truncated to order $n$. Each term $\hat{K}^{[i]}(t)$ scales as $O(\tilde{\Omega}^{-i})$, with the boundary condition $\int^{NT}_0dt \hat{K}^{[i]}(t)=0$ imposed in the van Vleck gauge~\cite{Bukov2015,Eckardt2015,Mikami2016}. This rotation yields a truncated static Hamiltonian $\hat{D}_n$ up to order $n\leq n_{*}=O(\tilde{\Omega})$~\cite{Mori2016,Kuwahara2016}, as given by
\begin{align} \label{eq:Rotation to the static Floquet Hamiltonian in the interaction picture}
    \hat{H}^{R}_{int}(t)&=\mathcal{\hat{U}}^{\dag}_n(t)\left [ \hat{H}_{int}(t)-i\partial_t \right ]\mathcal{\hat{U}}_n(t)\\ 
    &\simeq \hat{D}_n+\hat{V}^{[n]}_{int}(t)+O(\tilde{\Omega}^{-n-1}), \nonumber
\end{align}
where $\hat{D}_n=\sum^{n}_{i=0}\hat{V}^{[i]}_{vV}$, with $\hat{V}^{[i]}_{vV}$ scaling as $O(\tilde{\Omega}^{-i}\lambda)$. The residual time-dependent term in the dressed Hamiltonian $\hat{H}^{R}_{int}(t)$, $\hat{V}^{[n]}_{int}(t)=\partial_t K^{[n+1]}(t)$, is $NT$-periodic and of order $O(\tilde{\Omega}^{-n}\lambda)$. The evolution operator $\hat{U}_{int}(t_2,t_1)=\mathcal{\hat{U}}_n(t_2)\mathcal{T}[e^{-i\int_{t_1}^{t_2} dt\hat{H}^{R}_{int}(t)}]\mathcal{\hat{U}}^{\dag}_n(t_1)$, governed approximately by the prethermal Hamiltonian $\hat{D}_n$ in Eq.~\eqref{eq:Rotation to the static Floquet Hamiltonian in the interaction picture}, simplifies to 
\begin{equation} \label{eq:Time evolution of U_int}
    \hat{U}_{int}(t_2,t_1)\simeq \mathcal{\hat{U}}_n(t_2)e^{-i\hat{D}_n (t_2-t_1)}\mathcal{\hat{U}}^{\dag}_n(t_1),
\end{equation}
defining the prethermal regime for times exponentially long in $\tilde{\Omega}$, provided $1\ll \tilde{\Omega}$. %For stroboscopic evolution over $NT$ periods, the evolution operator becomes $\hat{U}(NT)=\hat{X}^{N}\hat{U}_{int}(NT)=\hat{U}_{int}(NT)\sim \mathcal{\hat{U}}_n(NT)e^{-i\hat{D}_n NT}\mathcal{\hat{U}}^{\dag}_n(0)=\mathcal{\hat{U}}_n(0)e^{-i\hat{D}_n NT}\mathcal{\hat{U}}^{\dag}_n(0)$ using the periodicity $\mathcal{\hat{U}}_n(t)=\mathcal{\hat{U}}_n(t+NT)$. 
Further details on the van Vleck expansion and explicit forms of the terms $\hat{V}^{[i]}_{vV}$ and $\hat{K}^{[i]}(t)$ are provided in the Supplemental Material (SM)~\cite{SM} (also see Refs.~\cite{Mikami2016,Mizuta2019,Mori2022}).

\section{Extended dynamical space-time symmetry group in the interaction picture}
Let us now derive the extended dynamical symmetry group, $G^{int}_{st}$, of the interaction picture Hamiltonian $\hat{H}_{int}(t)$, given the dynamical symmetry group $G_{st}=M\times A=\mathbb{Z}_2\times \mathbb{Z}^{T}_2$ of the original Hamiltonian $H(t)$. The element $\hat{X}$ in Eq.~\eqref{eq:Definition of Symmetry X} forms a subgroup $\mathbb{Z}_N\in G^{int}_{st}$ through the transformation
\begin{equation} \label{eq:Element X of G_int}
    \hat{X}^{m}\hat{H}_{int}(t)\hat{X}^{-m}=\hat{H}_{int}(t-mT).
\end{equation}
This emergent $\mathbb{Z}_N$ symmetry of $\hat{H}_{int}(t)$, absent in $\hat{H}(t)$, arises solely from the resonant drive $\hat{H}_0(t)$ and leads to prethermal DTCs~\cite{Else2020,Else2017_1,Mizuta2019}. %Here, $\hat{X}$ generates a time shift by multiples of the period $T$, echoing the discrete time translation symmetry embedded in the original Hamiltonian. 

For a unitary element $\hat{g}_M\in M\subset G_{st}$, defined in Eq.~\eqref{eq:Definition of symmetries}, the corresponding interaction picture element is identified as $\hat{g}^{int}_M=\hat{U}^{-1}_0(T/2)\hat{g}_M$, where $\hat{g}^{int}_M$ represents the projective realization of $\hat{g}_M$. This is demonstrated by the relation $[\hat{g}^{int}_M]^2=\hat{X}^{-1}\hat{g}^2_M$. Similarly, for an antiunitary element $\hat{g}_A\in A\subset G_{st}$, defined by setting $s=T/2$ in Eq.~\eqref{eq:Definition of symmetries}, the interaction picture counterpart is $\hat{g}^{int}_{A}=\hat{U}^{-1}_0(T/2)\hat{g}_A$ with the corresponding relation $[\hat{g}^{int}_A]^2=\hat{g}^2_A$. These elements act on $\hat{H}_{int}(t)$ according to symmetry relation
%unitary element transforms $\hat{H}_{int}(t)$ to $\hat{H}_{int}(t+T/2)$ while the antiunitary instead takes it to $\hat{H}_int(T/2-t)$.
%
\begin{equation} \label{eq:Element g_M,g_A of G_int}
    \hat{g}^{int}_{M/A}\hat{H}_{int}(t)[\hat{g}^{int}_{M/A}]^{-1}=\hat{H}_{int}(T/2\pm t).
\end{equation}
The algebraic structure of $G^{int}_{st}$ is further defined by %the commutativity of the unitary element $\hat{g}^{int}_M$ with $\hat{X}$, and the automorphism of the antiunitary element $\hat{g}^{int}_A$ with $\hat{X}$, given by
\begin{equation} \label{eq:Commute between g_M and X and g_A and X}
    \hat{g}^{int}_M \hat{X} [\hat{g}^{int}_M]^{-1}=\hat{X},\quad \hat{g}^{int}_A \hat{X} [\hat{g}^{int}_A]^{-1}=\hat{X}^{-1}.
\end{equation}
These relations establish the extended dynamical symmetry group in the interaction picture as $G^{int}_{st}=\mathbb{Z}_2 \times (\mathbb{Z}_N \rtimes \mathbb{Z}^{T}_2)$, encapsulating the interplay between the emergent $\mathbb{Z}_N$ symmetry and the original symmetry group $G_{st}$. Detailed derivations and further discussions on these symmetry elements are provided in the SM~\cite{SM}. 

\begin{table}[b]
\caption{\label{tab:Symmetry tables}%
Mapping dynamical space-time symmetries to static symmetries of the prethermal Hamiltonian.}
\begin{ruledtabular}
\begin{tabular}{lcc}
\multicolumn{1}{c}{}&
\multicolumn{1}{c}{$\hat{H}(t)$} & 
\multicolumn{1}{c}{$\hat{D}_n$\footnote{Prethermal Hamiltonian truncated to $n\leq n_{*}=O(\tilde{\Omega})$ in Eq.~\eqref{eq:Rotation to the static Floquet Hamiltonian in the interaction picture}}} \\
\hline
Symmetry Group & $G_{st}$ & $G^{int}_{st}$  \\
Symmetry Elements\footnote{$\hat{g}_{M,A}\in G_{st}$ in Eq.~\eqref{eq:Definition of symmetries}, $\hat{g}^{int}_{M/A}=\hat{U}^{-1}_0(T/2)\hat{g}_{M/A}$, $\hat{X}$ in Eq.~\eqref{eq:Definition of Symmetry X}} & $\hat{g}_M,~\hat{g}_A$ & $\hat{g}^{int}_M,~\hat{g}^{int}_A,~\hat{X}$ \\
Group Algebra & $\mathbb{Z}_2\times\mathbb{Z}^{T}_2$ & 
$\mathbb{Z}_2\times(\mathbb{Z}_N\rtimes\mathbb{Z}^{T}_2)$  \\
\end{tabular}
\end{ruledtabular}
\end{table}
\begin{figure}[t]
    \centering
    \includegraphics[width=\columnwidth]{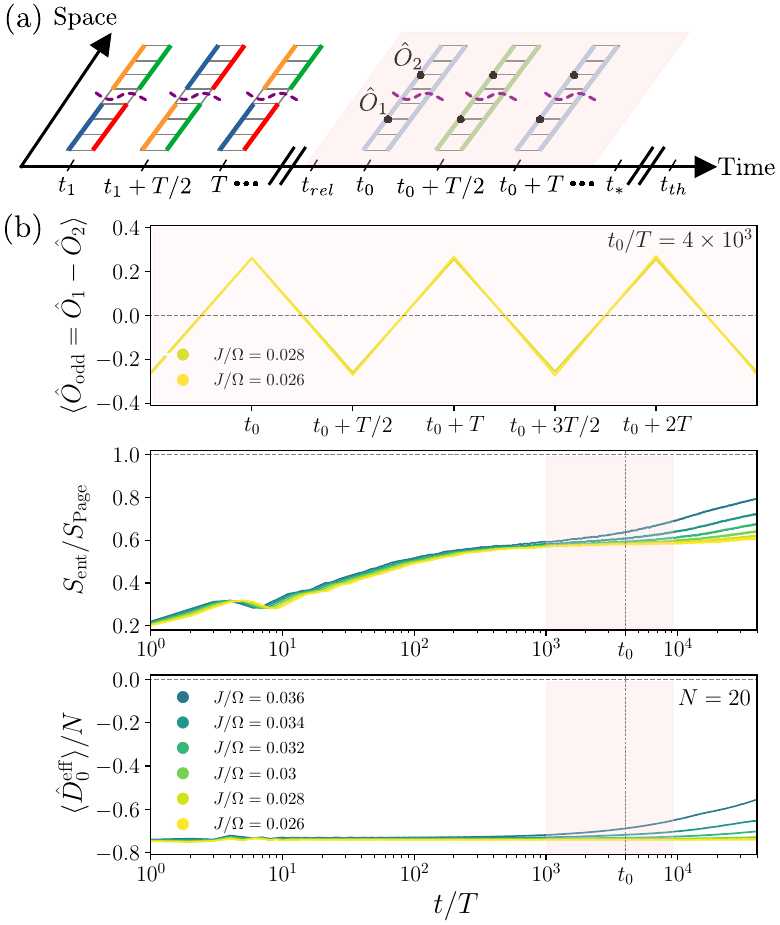}
    \caption{(a) Schematic illustrating the Floquet thermalization of a two-leg spin-$1/2$ ladder model with a Hamiltonian preserving the order-$2$ dynamical symmetry $\hat{g}_M$ in Eq.~\eqref{eq:Definition of symmetries}. The upper left and right legs are colored in orange and green, and the lower left and right legs are colored in blue and red. After relaxation at $t_{rel}$, the system enters a prethermal regime governed approximately by the static prethermal Hamiltonian in Eq.~\eqref{eq:Rotation to the static Floquet Hamiltonian in the interaction picture} until $t_{*}$%=n_{*}T$ ($n_{*}=O(\tilde{\Omega})$)
    , with $t_0$ marking a reference time within this regime. Beyond $t_{*}$, heating drives the system to an infinite temperature state at $t_{th}$. (b) Time evolution of $\langle \hat{O}_{\rm odd}(t)\rangle $ with $e^{i\alpha_M}=-1$ at $t=mT$ and $t=mT+T/2$ in the prethermal regime verifies Eq.~\eqref{eq:final expression comparing two values}, detecting $\hat{g}_M$. Shaded regions denote the prethermal regime, with bipartite entanglement and energy density of the prethermal Hamiltonian consistent across a range of frequencies. Model parameters in Eq.~\eqref{eq:H_0 and V}: $\tau=0.25$, $J_1=\Omega$, $J^{\prime}=J/(0.5-\tau)=4$, $\lambda^{LR}_a=\lambda^{LR}_b=0.5$, $g_x=g_y=0.45225$, $g_z=0.7$, and $g_{zz}=1.3$ (see the SM~\cite{SM} for details).}
    \label{fig:micromotion in prethermalization}
\end{figure}

\section{Symmetry group of the prethermal Hamiltonian}
Now we have established the dynamical symmetry group in the interaction picture, we turn to symmetry group of the truncated prethermal Hamiltonian $\hat{D}_n$. This inherits the symmetries $\hat{X}$, $\hat{g}^{int}_M$, and $\hat{g}^{int}_A$ from $G^{int}_{st}$ of $\hat{H}_{int}(t)$, as specified in Eqs.~\eqref{eq:Element X of G_int} and \eqref{eq:Element g_M,g_A of G_int}. We define $\hat{V}_m$ as 
\begin{equation} \label{eq:V_m}
    \hat{V}_m=\frac{1}{NT}\int^{NT}_0dt\hat{H}_{int}(t)e^{im(\Omega/N)t}.
\end{equation}
The $i$-th order van Vleck expansion term, $\hat{V}^{[i]}_{vV}$, in $\hat{D}_n$ contains terms with numerators of the form $\hat{V}_{m_1}\hat{V}_{m_2}\cdots \hat{V}_{m_i}$. Symmetry operations act on these terms as follows
\begin{align} \label{eq:Symmetry actions onto van Vleck expansion of D_n}
    &\hat{X}\hat{V}_{m_1}\cdots \hat{V}_{m_i}\hat{X}^{-1}=e^{i\frac{2\pi}{N}(\sum^{i}_{k=1}m_k)}\hat{V}_{m_1}\cdots \hat{V}_{m_i}, \\
    &\hat{g}^{int}_M\hat{V}_{m_1}\cdots \hat{V}_{m_i}[\hat{g}^{int}_M]^{-1}=e^{-i\frac{2\pi}{2N}(\sum^{i}_{k=1}m_k)}\hat{V}_{m_1}\cdots \hat{V}_{m_i}, \nonumber \\
    &\hat{g}^{int}_A\hat{V}_{m_1}\cdots \hat{V}_{m_i}[\hat{g}^{int}_A]^{-1}=e^{-i\frac{2\pi}{2N}(\sum^{i}_{k=1}m_k)}\hat{V}_{m_1}\cdots \hat{V}_{m_i}. \nonumber
\end{align}
The van Vleck gauge condition imposes a constraint on the indices, $\sum^{i}_{k=1}m_k=0$~\cite{Eckardt2015,Mikami2016,Mizuta2019}, leading to commutation relations
\begin{equation} \label{eq:Symmetries of prethermal Hamiltonian D_n}
    \left[ \hat{X}, \hat{D}_n \right]=\left[ \hat{g}^{int}_M, \hat{D}_n \right]=\left[ \hat{g}^{int}_A, \hat{D}_n\right]=0,
\end{equation}
valid for any integer $n\leq n_{*}=O(\tilde{\Omega})$. Thus, the prethermal Hamiltonian $\hat{D}_n$ inherits the full symmetry group $G^{int}_{st}$ of $\hat{H}_{int}(t)$, as specified in Table~\ref{tab:Symmetry tables}.

Finally, we emphasize that the rotation operator $\mathcal{\hat{U}}_n(t)$ in Eq.~\eqref{eq:Rotation to the static Floquet Hamiltonian in the interaction picture} also transforms under the symmetry group. In stroboscopic dynamics over $NT$ periods, it is often treated as static by setting $\mathcal{\hat{U}}_n(0)=\mathcal{\hat{U}}_n(NT)$, assuming commutation with the symmetry elements of Hamiltonian that has a static symmetry group~\cite{Else2017_1,Mizuta2019}. However, when the symmetry group includes dynamical elements, the transformations of $\mathcal{\hat{U}}_n(t)$ must be explicitly accounted for, as given by 
%the symmetry actions must be taken into account to derive the transformations for the rotation operator. The symmetry transformations of $\mathcal{\hat{U}}_n(t)$ are
%
\begin{align} \label{eq:Symmetry actions onto the rotation operator generated by the kick}
    \hat{X}^m\mathcal{\hat{U}}_n(t)\hat{X}^{-m}&=\mathcal{\hat{U}}_n(t-mT), \\ 
    \hat{g}^{int}_{M/A} \mathcal{\hat{U}}_n(t)[\hat{g}^{int}_{M/A}]^{-1}&=\mathcal{\hat{U}}_n(T/2\pm t). \nonumber 
\end{align}
Using Eqs.~\eqref{eq:Time evolution of U_int} and \eqref{eq:Symmetry actions onto the rotation operator generated by the kick}, the time evolution over $lT$ periods, with $l\bmod N$, is expressed as $\hat{U}(lT)=\hat{X}^{l}\hat{U}_{int}\simeq \hat{X}^{l}\mathcal{\hat{U}}_n(lT)e^{-i\hat{D}_n lT}\mathcal{\hat{U}}^{\dag}_n(0)=\mathcal{\hat{U}}_n(0)\hat{X}^le^{-i\hat{D}_n lT}\mathcal{\hat{U}}^{\dag}_n(0)$, which matches the expression in Ref.~\cite{Else2017_1}. Furthermore, we derive symmetry relations for $\hat{U}_{int}(t_2,t_1)$, given by $\hat{g}^{int}_{M/A}\hat{U}_{int}(t_2,t_1)[\hat{g}^{int}_{M/A}]^{-1}=\hat{U}_{int}(T/2\pm t_2,T/2\pm t_1)$~\cite{SM}. These relations are equivalent to Eq.~\eqref{eq:Element g_M,g_A of G_int}.  
%using the commutation relations in Eq.~\eqref{eq:Symmetries of prethermal Hamiltonian D_n}. 
%holding for the kick operator $\hat{K}_n(t)=\sum^{n}_{i=1}\hat{K}^{[i]}(t)$ transforms as $\hat{X}^m\hat{K}^{[i]}(t)\hat{X}^{-m}=\hat{K}^{[i]}(t-mT)$, $\hat{g}^{int}_M\hat{K}^{[i]}(t)[\hat{g}^{int}_M]^{-1}=\hat{K}^{[i]}(t+T/2)$, and $\hat{g}^{int}_A\hat{K}^{[i]}(t)[\hat{g}^{int}_A]^{-1}=-\hat{K}^{[i]}(T/2-t)$, leading to the following symmetry relations 
%
\begin{figure*}
    \includegraphics[width=\textwidth]{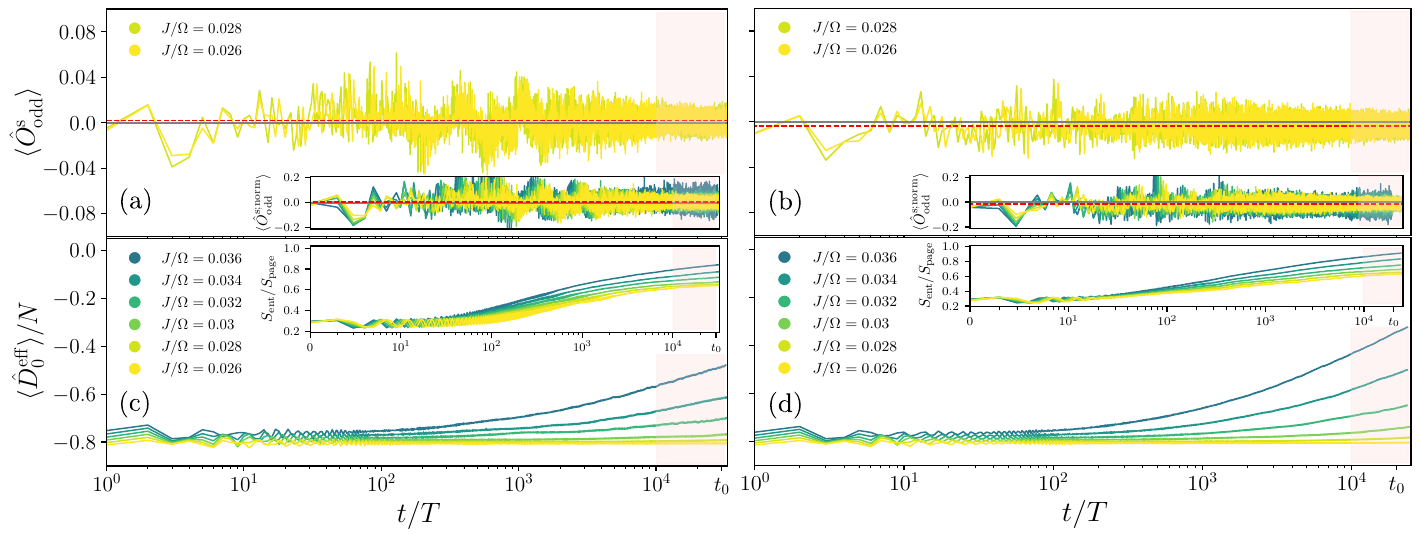}
    \caption{(a) Time evolution of $\hat{O}^{s}_{\rm odd}(mT)=\hat{O}_{\rm odd}(mT)+\hat{O}_{\rm odd}(mT+T/2)$ stabilizes at zero (red line) in the prethermal regime, confirming the presence of $\hat{g}_M$. (b) In the symmetry-broken case, $\hat{O}^{s}_{\rm odd}(mT)$ stabilizes at a non-zero value, indicating symmetry breaking. The red line marks the center of stable oscillations in the prethermal regime. (c) and (d) Energy density and bipartite entanglement entropy (insets) highlight the prethermal regime in shaded regions. The parameters used match Fig.~\ref{fig:micromotion in prethermalization}, except for the symmetry-preserving $(\lambda^{LR}_a,\lambda^{LR}_b)=(1,1)$ and symmetry-broken $(\lambda^{LR}_a,\lambda^{LR}_b)=(0.8,1.2)$ cases. }
    \label{fig:Comparison with and without symmtry}
\end{figure*} 

\section{Engineering prethermal micromotion with dynamical symmetries}
We propose a scheme to engineer nontrivial micromotion in the prethermal stage using dynamical symmetries in $G_{st}$, focusing on the order-$2$ unitary symmetry $\hat{g}_M\in G_{st}$ of $\hat{H}(t)$ in Eq.~\eqref{eq:Definition of symmetries}, which can be generalized to higher-order symmetries. For a set of Hermitian operators $\hat{O}$ satisfying the conditions
\begin{equation} \label{eq:set of local operators O}
    \hat{g}_M\hat{O}\hat{g}^{-1}_M=e^{i\alpha_M}\hat{O},\quad[\hat{g}^{int}_M,\hat{O}]=[\hat{D}_0,\hat{O}]=0,
\end{equation}
where $e^{i\alpha_M}=\pm1$ due to Hermiticity, we examine the expectations values $\langle\hat{O}(mT)\rangle$ and %=\Tr[\hat{\rho}(mT)\hat{O}_i]$ 
$\langle\hat{O}(mT+T/2)\rangle$ for integer $m$ in the prethermal regime. Using the projective symmetry $\hat{g}^{\prime}_M=\hat{U}^{-1}(T/2)\hat{g}_M$ of the extended symmetry group for the Floquet operator $\hat{U}(T)$~\cite{Na2023}, with $\hat{g}^{\prime 2}_M=\hat{U}^{-1}(T)\hat{g}^2_M$, the expectation value at $t=mT+T/2$ can be expressed as  
\begin{equation} \label{eq:expectation value of O at mT+T/2}
    \langle \hat{O}(mT+T/2)\rangle=e^{-i\alpha_M}\Tr\left [ \hat{g}^{\prime -1}_M\hat{\rho}(mT)\hat{g}^{\prime}_M\hat{O}\right],
\end{equation}
where $\hat{\rho}(t)$ is the density matrix of the system at time $t$. By applying $\hat{g}^{\prime -1}_M\mathcal{\hat{U}}_n(0)=\mathcal{\hat{U}}_n(0)e^{-i\hat{D}_n T/2}[\hat{g}^{int}_M]^{-1}$ (derived in SM~\cite{SM}) and $\hat{U}(mT)\simeq \mathcal{\hat{U}}_n(0)\hat{X}^me^{-i\hat{D}_n mT}\mathcal{\hat{U}}^{\dag}_n(0)$, the term $\hat{g}^{\prime -1}_M\hat{\rho}(mT)\hat{g}^{\prime}_M$ simplifies to $\hat{A}\hat{\rho}(0)\hat{A}^{-1}$, where
\begin{equation} \label{eq:operator A}
    \hat{A}=\mathcal{\hat{U}}_n(0)e^{-i\hat{D}_n T/2}[\hat{g}^{int}_M]^{-1}\hat{X}^me^{-i\hat{D}_n mT}\mathcal{\hat{U}}^{\dag}_n(0).
\end{equation}
Approximating $\mathcal{\hat{U}}_n(0)\sim\mathbb{I}$, we obtain $\langle \hat{O}(mT+T/2)\rangle \simeq e^{-i\alpha_M}\Tr[\hat{X}^m e^{-i\hat{D}_n mT}\hat{\rho}(0) e^{i\hat{D}_n mT}\hat{X}^{-m}\hat{B}]$, where $\hat{B}=\hat{g}^{int}_Me^{i\hat{D}_nT/2}\hat{O}e^{-i\hat{D}_nT/2}[\hat{g}^{int}_M]^{-1}$. For comparison, the expectation value at $t=mT$ is $\langle \hat{O}(mT)\rangle\simeq \Tr[\hat{X}^m e^{-i\hat{D}_n mT}\hat{\rho}(0)e^{i\hat{D}_n mT}\hat{X}^{-m}\hat{O}]$. Using $[\hat{g}^{int}_M,\hat{O}]=0$ from Eq.~\eqref{eq:set of local operators O} and the approximation $[\hat{D}_n,\hat{O}]\simeq0$ valid for $1\ll \tilde{\Omega}$, we find $\hat{B}\simeq \mathbb{I}$, leading to the following relation between the two expressions
\begin{equation} \label{eq:final expression comparing two values}
    \langle \hat{O}(mT+T/2)\rangle \simeq e^{-i\alpha_M}\langle \hat{O}(mT)\rangle.
\end{equation}
\section{Numerical simulation}
As a concrete example, let us consider a two-leg, spin-$1/2$ ladder of length $L$ [Fig.~\ref{fig:micromotion in prethermalization}(a)] under dual driving, with an order-2 symmetry, $\hat{g}_M\in G_{st}=\mathbb{Z}_2$. In particular, $\hat{g}_M$ is a mirror operation along the centerline of the ladder that maps the left and right halves of the ladder onto one other, as shown in Fig.~\ref{fig:micromotion in prethermalization}(a). To enforce this symmetry in the dual-driving setup, we implement a four-step driven model,
\begin{equation} \label{eq:four-step driven model}
    \hat{H}(t)=\begin{cases}
         \hat{H}_{0,a}&(0 \leq t < \tau T), \\ 
         \hat{V}_{a}+\hat{V}^{sc}_{a}&(\tau T \leq t < T/2), \\ 
         \hat{H}_{0,b}&(T/2 \leq t < \tau T+T/2),\\ 
         \hat{V}_{b}+\hat{V}^{sc}_{b}&(\tau T+T/2 \leq t < T), 
         \end{cases}
\end{equation}
satisfying the symmetry relations
$\hat{g}_M\hat{H}_{0,a}\hat{g}^{-1}_M=\hat{H}_{0,b}$ and $\hat{g}_M\hat{V}_{a}\hat{g}^{-1}_M=\hat{V}_{b}$. The resonant $\hat{H}_{0,a/b}$ and the weak drive $\hat{V}_{a/b}$, we choose  
\begin{align} \label{eq:H_0 and V}
    \hat{H}_{0,a}=&J_1[\sum^{L/2-2}_{i=0}(S^z_iS^z_{i+1}+\sigma^x_i\sigma^x_{i+1})+\sum^{L-2}_{i=L/2}(S^y_iS^y_{i+1}+\sigma^z_i\sigma^z_{i+1})],  \nonumber \\
    \hat{V}_a=&J^{\prime}[\sum^{L/2-2}_{i=0}(S^x_iS^x_{i+1}+\sigma^y_i\sigma^y_{i+1})+\sum^{L/2-2}_{i=0} g_{zz}S^z_i\sigma^z_i \nonumber \\
    &+\sum^{L/2-2}_{i=0}(g_{x}S^x_i+g_{z}S^z_i+g_{y}\sigma^y_i+g_{z}\sigma^z_i)],  
\end{align}
where the spins in the upper (lower) chain are labeled by $S~(\sigma)$. Here, we choose $J_1=\Omega/4\tau$ to ensure that the resonant drive $\hat{H}_{0}(t)$ generates the $\mathbb{Z}_2$ symmetry operation $\hat{X}=e^{-i\hat{H}_{0,b}\tau T}e^{-i\hat{H}_{0,a}\tau T}$, where $\hat{X}^2=\mathbb{I}$.

Finally, we introduce an additional coupling $\hat{V}_{a/b}^{sc}$ at the center of the ladder
\begin{align} \label{eq:V sym breaking}
&\hat{V}^{sc}_{a/b}=J^{\prime}\lambda^{LR}_{a/b}[S^{x}_{*}S^{x}_{*+1}+\sigma^y_{*}\sigma^y_{*+1}+g_{zz}S^z_{*+1/*}\sigma^z_{*+1/*} \nonumber \\
&+g_{x}S^x_{*+1/*}+g_{z}S^z_{*+1/*}+g_{y}\sigma^y_{*+1/*}+g_{z}\sigma^z_{*+1/*}], 
\end{align}
where $*=L/2-1$. 
This allows us to break the symmetry by setting $\lambda^{LR}_{a}\neq \lambda^{LR}_{b}$, causing $\hat{g}_M\hat{V}^{sc}_a\hat{g}^{-1}_M\neq\hat{V}^{sc}_b$. 
This Hamiltonian can be experimentally implemented using platforms such as superconducting quantum processor~\cite{Sameti2019} and cold atoms~\cite{Atala2014,Sompet2022}, providing a concrete avenue for its experimental realization.

To characterize Floquet thermalization dynamics~\cite{Machado2019}, we monitor energy absorption by the energy density of the effective prethermal Hamiltonian, $\langle \hat{D}^{\rm eff}_0(t)\rangle/N$, where $N=2L=20$ is the total number of spins. % and the initial state is arbitrary not necessarily an eigenstate of $\hat{D}^{\rm eff}_0$~\cite{SM}. 
Furthermore, we track the evolution of the bipartite entanglement entropy, $S_{\rm ent}(t)=-\Tr[\hat{\rho}_{L/2}\ln{\hat{\rho}_{L/2}}]$, where $\hat{\rho}_{L/2}=\Tr_{LH}[\ket{\Psi(t)}\bra{\Psi(t)}]$, normalized by the Page value, $S_{\rm page}=(L\ln{2}-1)/2$~\cite{Page1993}, signifying full thermalization to a featureless infinite temperature state. Initially, $S_{\rm ent}(t)$ grows during relaxation up to a time $t_{rel}$, after which it reaches a prethermal plateau that persists until $t_{*}\sim n_{*}T$, where $n_{*}=O(\tilde{\Omega})$. Beyond $t_{*}$, the system heats towards an infinite-temperature state, consistent with energy conservation during the prethermal regime and subsequent heating for a wide range of frequencies, as illustrated in Fig.~\ref{fig:micromotion in prethermalization}(b).

To probe the dynamical symmetry, we introduce local operators $\hat{O}_1=S^{x}_{L/2-1}$ and $\hat{O}_2=S^{x}_{L/2}$, defining the odd operator $\hat{O}_{\rm odd}=\hat{O}_1-\hat{O}_2$, which satisfies $\hat{g}_M \hat{O}_{\rm odd}\hat{g}^{-1}_M=-\hat{O}_{\rm odd}$, with $e^{i\alpha_M}=-1$ in Eq.~\eqref{eq:set of local operators O}. We verify the expected relation $\langle \hat{O}_{\rm odd}(mT+T/2)\rangle=-\langle \hat{O}_{\rm odd}(mT)\rangle$ in Eq.~\eqref{eq:final expression comparing two values} through the time evolution of these operators. This confirms the emergence of nontrivial micromotion induced by the dynamical symmetry $\hat{g}_M$, which extends beyond the conventional stroboscopic views of prethermal Floquet phases. As shown in  Fig.~\ref{fig:micromotion in prethermalization}(b), the prethermal regime is indicated by a shaded region, with $t_0$ serving as a reference temporal point within this regime.

\section{Sensitivity to dynamical symmetry breaking}
To verify the perfect oscillation of $\braket{\hat{O}_{\rm odd}(mT)} = -\braket{\hat{O}_{\rm odd}(mT+T/2)}$ is directly tied to the dynamical symmetry, we explicitly break the symmetry $\hat{g}_M$, by adjusting the coupling across the center bond by setting $\lambda^{LR}_a\neq \lambda^{LR}_b$ in Eq.~\eqref{eq:V sym breaking}. We define $\hat{O}^{s}_{\rm odd}(mT)=\hat{O}_{\rm odd}(mT)+\hat{O}_{\rm odd}(mT+T/2)$ for integer $m$, which satisfies $\hat{O}^{s}_{\rm odd}(mT)=0$ when the symmetry $\hat{g}_M$ is preserved, and during the prethermal regime $t_{rel}\leq mT\leq t_{*}$. Following relaxation, as the system enters the prethermal regime, $\hat{O}^{s}_{\rm odd}(t)$ saturates to zero when the symmetry is maintained, as shown in Fig.~\ref{fig:Comparison with and without symmtry}(a) for $(\lambda^{LR}_{a},\lambda^{LR}_{b})=(1,1)$. In contrast, when the symmetry is broken by setting $(\lambda^{LR}_a,\lambda^{LR}_b)=(0.8,1.2)$, the saturated value of $\hat{O}^{s}_{\rm odd}(t)$ is nonzero, as indicated by the location of the red horizontal line in Fig.~\ref{fig:Comparison with and without symmtry}(b). 

At infinite temperature, $\hat{O}_{\rm odd}(mT)$ and $\hat{O}_{\rm odd}(mT+T/2)$ both approach zero, obscuring whether $\hat{O}^{s}_{\rm odd}(t)=0$ genuinely reflects the symmetry preservation or simply the independent vanishing of each term. To resolve this ambiguity, we introduce the normalized quantity $\hat{O}^{s;\rm norm}_{\rm odd}(mT)=\hat{O}^{s}_{\rm odd}(mT)/||\hat{O}_{\rm odd}(mT)||$, shown in the insets of Fig.~\ref{fig:Comparison with and without symmtry}(a) and (b), which distinguishes symmetry-preserving prethermal dynamics from thermalizing states approaching infinite temperature. 

\section{Conclusion and Outlook}
We investigate generic Floquet prethermalization under coexisting resonant and high-frequency drives, providing a framework to identify the symmetry group of the prethermal Hamiltonian and its algebra based on the dynamical space-time symmetry group of the driven system. Unlike previous studies that focused on static symmetries, our approach incorporates dynamical symmetries, with the kick operator playing a central role in the formulation. We validate this framework using a spin-ladder model with an order-two dynamical symmetry and confirm the symmetry-predicted relations between local observables at distinct temporal points within the Floquet period, revealing nontrivial micromotion that extends beyond conventional stroboscopic Floquet dynamics. Importantly, this phenomenon is independent of the initial state, enhancing its experimental feasibility.

This work opens two promising directions for future exploration. First, the spontaneous symmetry breaking of projected static symmetries in the prethermal Hamiltonian could be extended to study discrete space-time crystalline orders, generalizing previous research on prethermal time-crystalline phases~\cite{Else2020,Zaletel2023}. Moreover, the identification of the extended symmetry group in the prethermal regime could be applied to quasiperiodically driven systems~\cite{Else2020_2,He2023,Gallone2024}, revealing richer group algebras and allowing the discovery of novel phases characterized by enriched spatiotemporal orders.

\begin{acknowledgments}
I. N. thanks M. Bukov, P. Crowley, F. Machado, A. Polkovnikov, R-J. Slager, and K. Takasan for stimulating discussions on related projects. We used DYNAMITE~\cite{GregDMeyer2024} to simulate the dynamics of the quantum systems. This work (I.N, S.M.G.) was supported by the Theory of Materials Program (KC2301) funded by the US Department of Energy, Office of Science, Basic Energy Sciences, Materials Sciences and Engineering Division under Contract No. DE-AC02-05CH11231. The work performed at the Molecular Foundry was supported by the Office of Science, Office of Basic Energy Sciences, of the US Department of Energy under the same contract No. Computational resources were provided by the National Energy
Research Scientific Computing Center (NERSC). This work was supported in part by the US Army Research Office (grant no. W911NF-24-1-0079). Y.P. is supported by the NSF PREP grant (PHY-2216774). 
\end{acknowledgments}

%\clearpage
% ---- Smooth spacing before Supplementary ----
%\vspace*{2em} % tweak if you want slightly more or less

% Switch to "S" numbering for SI
\renewcommand{\thefigure}{S\arabic{figure}}
\renewcommand{\thetable}{S\arabic{table}}
\renewcommand{\theequation}{S\arabic{equation}}
\renewcommand{\thesection}{S\Roman{section}}

\setcounter{section}{0}
\setcounter{figure}{0}
\setcounter{table}{0}
\setcounter{equation}{0}

\makeatletter
% Make all subsections in the SI unnumbered (no A., B., etc.)
\let\oldsubsection\subsection
\renewcommand{\subsection}{\oldsubsection*}

\let\oldsubsubsection\subsubsection
\renewcommand{\subsubsection}{\oldsubsubsection*}
\makeatother

%\onecolumngrid % OPTIONAL: comment out if you want SI strictly two-column
%\twocolumngrid % keep two-column; leave as-is for pure two-column SI
%\begin{widetext}
\section*{Supplemental Material}

\section{Prethermalization in a dual-driving}

This section discusses the Floquet prethermalization and derives the prethermal Hamiltonian in the presence of coexisting resonant and high-frequency drives. 

We consider a spin system where each spin resides in a $d$-dimensional Hilbert space. The spins are indexed by $i=1,2,\cdots,N_s$, with the set of all spins denoted by $\Lambda$. The Hamiltonian is given by $\hat{H}(t)=\hat{H}_0(t)+\hat{V}(t)$, where both $\hat{H}_0(t)$ and $\hat{V}(t)$ are time-periodic with the same period $T$. Here, $\hat{H}_0(t)$ is treated as the resonant drive, and $\hat{V}(t)$ as the high-frequency drive. The Floquet operator associated with the resonant drive $\hat{H}_0(t)$ is defined as
\begin{equation} \label{eq:SI Symmetry X def}
    \hat{X} = \hat{U}_0(T,0) = \mathcal{T}[e^{-i\int_{0}^{T} dt\hat{H}_0(t)}],\quad\hat{X}^N=\mathbb{I},
\end{equation}
which implements a spatially local $\mathbb{Z}_N$ unitary action. 

\subsection{$k$-locality and $J$-extensiveness constraints}
We constrain the drive such that the Hamiltonian contains at most $k$-body interactions, where $k$ is finite. This is expressed as 
\begin{equation} \label{eq:SI Spatially local V(t)}
    \hat{H}_0(t)=\sum_{|M|\leq k}\hat{h}_{M}(t),\quad \hat{V}(t)=\sum_{|M|\leq k}\hat{v}_{M}(t),
\end{equation}
where $\hat{h}_M(t)$ and $\hat{v}_{M}(t)$ are operators acting on subsets $M$ of the system, and $|M|$ denotes the number of sites in the subset $M$. Additionally, we impose the condition that the terms in the resonant drive commute among different subsets $M$ and $N$ as $[\hat{h}_M(t),\hat{h}_N(t)]=0$.  

The local interaction strength $J$ of the high-frequency drive is defined as 
\begin{equation} \label{eq:SI Local energy scale of V(t)}
    \sum_{M:i\in M}\left \| \hat{v}_{M}(t) \right \|\leq J\quad\text{for}\quad\forall i\in\Lambda=\left \{1,2,\cdots,N_s\right \},
\end{equation}
where the summation is over all subsets $M$ of the system containing the spin at site $i$. The typical energy scale of the weak drive $\hat{V}(t)$ is characterized by $\lambda\sim2kJ\ll \Omega$, where $\Omega=2\pi/T$ is the driving frequency. 

\subsection{Transition to the interaction picture}
To examine prethermalization under this dual-driving setup, we transition to the interaction picture with respect to $\hat{H}_0(t)$. The interaction picture Hamiltonian is given as 
\begin{equation} \label{eq:SI Interaction picture Hamiltonian}
    \hat{H}_{int}(t)=i[\partial_t\hat{U}^{\dag}_0(t)]\hat{U}_0(t)+\hat{U}^{\dag}_0(t)\hat{H}(t)\hat{U}_0(t).
\end{equation}
It follows that $\hat{H}_{int}(t)=\hat{U}^{\dag}_0(t)\hat{V}(t)\hat{U}_0(t)=\hat{V}_{int}(t)$, which is $NT$-periodic since $\hat{U}_0(NT)=\hat{X}^N=\mathbb{I}$ in Eq.~\eqref{eq:SI Symmetry X def}. The time evolution for one period $NT$, governed by $\hat{H}_{int}(t)$, is expressed as $\hat{U}_{int}(NT)=\hat{U}^{-1}_0(NT)\hat{U}(NT)=\hat{U}(NT)$. The Floquet Hamiltonian $\hat{V}^{FM}_{int}$ in the interaction picture is defined through 
\begin{equation} \label{eq:SI Interaction picture Floquet Hamiltonian}
    \hat{U}_{int}(NT)=\mathcal{T}[e^{-i\int^{NT}_{0}dt \hat{V}_{int}(t)}]=e^{-i\hat{V}^{FM}_{int}NT}. 
\end{equation}

As $\hat{V}(t)$ includes at most $k$-body interactions, as specified in Eq.~\eqref{eq:SI Spatially local V(t)}, and the terms in $\hat{H}_0(t)$ commute, the interaction picture transformation via $\hat{U}_0(t)$ ensures that $\hat{V}_{int}(t)$ retains at most $\tilde{k}$-body interactions, maintaining the finiteness of $\tilde{k}\sim k$. Additionally, the energy at each site of $\hat{V}_{int}(t)$ is also bounded by $\tilde{J}$, with $\sum_{M:i\in M}\left \| \hat{v}_{int,M}(t) \right \|\leq \tilde{J}\sim J$. We define
\begin{align} \label{eq:SI V_0, lambda, and order}
    &\hat{V}_0:=\sum_{|M|\leq \tilde{k}}\frac{1}{NT}\int^{NT}_0 dt \left \| \hat{v}_{int,M}(t) \right \|, \\
    &O(\tilde{\Omega}^n)=\textrm{const}\cdot(\frac{\Omega}{N\lambda_{int}})^n,\quad O(\tilde{T}^n)=\textrm{const}\cdot(NT\lambda_{int})^n, \nonumber
\end{align}
where $\lambda_{int}=2\tilde{k}\tilde{J}\sim \lambda$ characterizes the typical properties of the weak $NT$-periodic Floquet system governed by $\hat{V}_{int}(t)$, assuming the high-frequency limit in the interaction picture ($\tilde{T}\ll 1$ or $1\ll\tilde{\Omega}$). The driving amplitude over one period is given by $\hat{V}_0NT$. 

\subsection{Derivation of the prethermal Hamiltonian}
Assuming $\lambda_{int}NT<1/4$, the theorems in Ref.~\cite{Kuwahara2016} can be applied. The stroboscopic dynamics can be approximated using the truncated Floquet-Magnus (FM) expansion, given by $\hat{V}^{FM,n}_{int}=\sum^{n}_{m=0}\hat{\Omega}_m(NT)^m$, where each term satisfies the bound
\begin{equation} \label{eq:SI Norm of V_FM interaction picture}
    \left \|\hat{\Omega}_m\right \| \leq\frac{2\hat{V}_0\lambda_{int}^m}{(m+1)^2}m!=:\bar{\Omega}_m,
\end{equation}
with $\hat{V}_0$ defined in Eq.~\eqref{eq:SI V_0, lambda, and order}. The error in approximating $\hat{U}_{int}(NT)=\hat{U}(NT)$ by $e^{-iV^{FM}_{int}NT}$ is bounded by
\begin{align} \label{eq:SI Floquet_Magnus expansion of V_FM interaction picture}
    &\left \| \hat{U}(NT)-e^{-iV^{FM,n}_{int}NT} \right \| \leq 6 \hat{V}_0 NT2^{-n_0}+\bar{\Omega}_{n+1}(NT)^{n+2} \nonumber \\
    &\leq\frac{3N_s}{2\tilde{k}}2^{-n_0}+\frac{2N_s}{\tilde{k}}[(n+1)\lambda_{int} NT]^{n+2} \nonumber \\
    &\simeq e^{-O(\tilde{\Omega})}+O(\tilde{\Omega}^{-n-2}).
\end{align}
Here, $n\leq n_0:=[1/(16\lambda_{int}NT)]=O(\tilde{\Omega})$, where $[\cdot]$ is the floor function, and $N_s$ is the total number of spins. The derivations uses the following relations: $\hat{V}_0\leq2\tilde{J}N_s=2(\tilde{J}\tilde{k})\cdot(N_s/\tilde{k})$, $\lambda_{int}NT\leq1/4$, and $(m+1)!/(m+2)^2\leq(m+1)^{m+2}$. These ensure the first two inequalities in Eq.~\eqref{eq:SI Floquet_Magnus expansion of V_FM interaction picture}.

Next, we apply a unitary rotation to shift from the FM gauge to the van Vleck (vV) gauge, as described in~\cite{Mikami2016} and further discussed in~\cite{Mizuta2019}. When truncating both high-frequency expansions to a finite order $n\leq n_0$, the two gauges are related as
\begin{align} \label{eq:SI Connecting FM and VV gauges}
    &\left \| e^{-i\hat{K}_n(NT)}e^{-i\hat{V}^{vV,n}_{int} NT}e^{i\hat{K}_n(0)}-e^{-iV^{FM,n}_{int}\cdot NT}\right \| \\
    &\leq \frac{16N_s}{3\tilde{k}}[4(n+1)\lambda_{int}NT]^{n+2}\simeq O(\tilde{\Omega}^{-n-2}), \nonumber
\end{align}
where $\hat{K}_n(NT)=\hat{K}_n(0)$ is the $NT$-periodic truncated kick operator up to the $n$-th order, with $\hat{K}_n(t)=\sum^{n}_{i=1}\hat{K}^{[i]}(t)$. Each term $\hat{K}^{[i]}(t)$ scales as $O(\tilde{\Omega}^{-i})$, and the boundary condition $\int^{NT}_0dt\hat{K
}^{[i]}(t)=0$ is imposed in the vV gauge. Additionally, $\hat{V}^{vV,n}_{int}=\sum^{n}_{i=0}\hat{V}^{[i]}_{vV}$, with $\hat{V}^{[i]}_{vV}$ scaling as $O(\tilde{\Omega}^{-i}\lambda_{int})$, represents the $n$-th order truncated vV expansion. This corresponds to the $n$-th order effective prethermal Hamiltonian $\hat{D}_n$ in Ref.~\cite{Else2017_1}. Accordingly, the error in approximating $\hat{U}(NT)$ by $e^{-i\hat{K}_n(NT)}e^{-i\hat{V}^{vV,n}_{int} NT}e^{i\hat{K}_n(0)}$ is 
\begin{align} \label{eq:SI Error between U(NT) and NT evolution with V_VV}
    &\left \| \hat{U}(NT)-e^{-i\hat{K}_n(NT)}e^{-i\hat{V}^{vV,n}_{int} NT}e^{i\hat{K}_n(0)}\right \| \\
    &\leq \left \| \textrm{Eq}.~\eqref{eq:SI Floquet_Magnus expansion of V_FM interaction picture} \right \| + \left \| \textrm{Eq}.~\eqref{eq:SI Connecting FM and VV gauges} \right \| \simeq e^{-O(\tilde{\Omega})}+O(\tilde{\Omega}^{-n-2}). \nonumber
\end{align}
In the limit $1\ll\tilde{\Omega}$, this validates that the stroboscopic time evolution over $NT$ periods is well approximated by the effective $n$-th order truncated prethermal Hamiltonian $\hat{V}^{vV,n}_{int}=\hat{D}_n$. 

\subsection{Van Vleck high-frequency expansion}
A rotated frame, known as the Floquet reference frame as described in Ref.~\cite{Mori2022}, connects $\hat{H}_{int}(t)$ to $\hat{V}^{FM}_{int}$. This connection is established via the transformation $\hat{V}^{FM}_{int}=\mathcal{\hat{U}}^{\dag}(t)[\hat{H}_{int}(t)-i\partial_t]\mathcal{\hat{U}}(t)$, where $\mathcal{\hat{U}}(t)=e^{-i\hat{K}(t)}$. In the main text, we truncate the rotation $\mathcal{\hat{U}}_n(t)$ to order $n$ and introduce the dressed Hamiltonian $\hat{H}^{R}_{int}(t)$, given by
\begin{align} \label{eq:SI Rotation to the static Floquet Hamiltonian in the interaction picture}
    \hat{H}^{R}_{int}(t)&=e^{i\hat{K}_n(t)}\left [ \hat{H}_{int}(t)-i\partial_t \right ]e^{-i\hat{K}_n(t)}\\ 
    &\simeq \hat{D}_n+\hat{V}^{[n]}_{int}(t)+O(\tilde{\Omega}^{-n-1}), \nonumber
\end{align}
where $V^{[n]}_{int}(t)=\partial_t\hat{K}^{[n+1]}(t)$ is $NT$-periodic and satisfies $\int^{NT}_0dt\hat{V}_{int}(t)=0$. We present the $n$-th order truncation of the high-frequency expansions for the van Vleck (vV) effective Hamiltonian, $\hat{D}_n=\hat{V}^{vV,n}_{int}=\sum^{n}_{i=0}\hat{V}^{[i]}_{vV}$, and the kick operator, $\hat{K}_n(t)=\sum^n_{i=1}\hat{K}^{[i]}(t)$, for $n\leq2$, as follows~\cite{Mikami2016,Mizuta2019,Mori2022}
\begin{align} \label{eq:SI vV expansions}
    \hat{V}^{[0]}_{vV}=&\hat{V}_0, \\
    \hat{V}^{[1]}_{vV}=&\sum_{m\neq0}\frac{[\hat{V}_{-m},\hat{V}_m]}{2m(\Omega/N)}, \nonumber \\
    \hat{V}^{[2]}_{vV}=&\sum_{m\neq0}\frac{[[\hat{V}_{-m},\hat{V}_0],\hat{V}_m]}{2m^2(\Omega/N)^2}+\sum_{m\neq0}\sum_{n\neq0,m}\frac{[[\hat{V}_{-m},\hat{V}_{m-n}],\hat{V}_n]}{3mn(\Omega/N)^2}, \nonumber \\
    \hat{K}^{[1]}(t)=&i\sum_{m\neq0}\frac{\hat{V}_m}{m(\Omega/N)}e^{-im(\Omega/N)t}, \nonumber \\
    \hat{K}^{[2]}(t)=&-i\sum_{m\neq0}\frac{[\hat{V}_m,\hat{V}_0]}{m^2(\Omega/N)^2}e^{-im(\Omega/N)t} \nonumber \\
    &-i\sum_{m\neq0}\sum_{n\neq0,m}\frac{[\hat{V}_n,\hat{V}_{m-n}]}{2mn(\Omega/N)^2}e^{-im(\Omega/N)t}\nonumber.
\end{align}
Here, $\hat{V}_m$ is defined as
\begin{equation} \label{eq: SI V_m}
    \hat{V}_m=\frac{1}{NT}\int^{NT}_0dt\hat{H}_{int}(t)e^{im(\Omega/N)t}.
\end{equation}
\section{Symmetry of prethermal Hamiltonian}
In this section, we derive the symmetry group of the prethermal Hamiltonian $\hat{D}_n$ and identify its group algebra, starting from the dynamical space-time symmetry group $G_{st}$ of the original Hamiltonian $\hat{H}(t)$. We first establish the extended dynamical space-time symmetry group $G^{int}_{st}$ of the interaction picture Hamiltonian $\hat{H}_{int}(t)$, and then identify the symmetry group of the truncated prethermal Hamiltonian $\hat{D}_n$.

\subsection{Dynamical symmetry group $G_{st}$}
We consider Floquet systems that exhibit the symmetry group $G_{st}$, encompassing both static and dynamical symmetries transforming both spatial and temporal coordinates. Within $G_{st}$, subgroups $M$ and $A$ represent unitary and antiunitary symmetries, respectively. For order-2 discrete symmetries, under the assumption that $\hat{g}_M\in M$ and $\hat{g}_A \in A$ commute, we have
\begin{align} \label{eq:SI Definition of symmetries}
    \hat{g}_M\hat{H}(t)\hat{g}^{-1}_M&=\hat{H}(t+T/2),  \\
    \hat{g}_A\hat{H}(t)\hat{g}^{-1}_A&=\hat{H}(-t+s), \nonumber
\end{align}
where $s\in\left[0,T\right]$ is a reference point within the driving period $T$. Without loss of generality, we set $s=T/2$.

\subsection{Extended dynamical symmetry group $G^{int}_{st}$}
The element $\hat{X}$, with $\hat{X}^N=\mathbb{I}$ in Eq.~\eqref{eq:SI Symmetry X def}, forms a subgroup $\mathbb{Z}_N\subset G^{int}_{st}$, with transformations
\begin{align} \label{eq:SI X_Subgroup of G_int}
    \hat{X}^{m}\hat{H}_{int}(t)\hat{X}^{-m}&=\hat{H}_{int}(t-mT), \\
    \hat{X}^{m}\hat{U}_{int}(t,0)\hat{X}^{-m}&=\hat{U}_{int}(t-mT,-mT). \nonumber
\end{align}
This emergent $\mathbb{Z}_N$ symmetry of $\hat{H}_{int}(t)$, absent in $\hat{H}(t)$, arises solely from the resonant drive $\hat{H}_0(t)$ and underpins the realization of prethermal discrete time crystals~\cite{Else2017_1,Mizuta2019}.

For a unitary element $\hat{g}_M\in M\subset G_{st}$, as defined in Eq.~\eqref{eq:SI Definition of symmetries} and $\hat{U}_{int}(t,0)=\mathcal{T}[e^{-i\int^t_0\hat{H}_{int}(t)}]=\hat{U}^{-1}_0(t)\hat{U}(t)$, the action of $\hat{g}_M$ is expressed as
\begin{align} \label{eq:SI g_M relations}
    \hat{g}_M\hat{U}_{int}(t)\hat{g}_M^{-1}&=\hat{U}_0(T/2,t+T/2)\hat{U}(t+T/2,T/2) \\
    &=\hat{U}_0(T/2)\hat{U}_{int}(t+T/2)\hat{U}^{-1}(T/2). \nonumber
\end{align}
We define the unitary symmetry element $\hat{g}^{int}_M\in G^{int}_{st}$ as $\hat{g}^{int}_M=\hat{U}^{-1}_0(T/2)\hat{g}_M$, which satisfies
\begin{align} \label{eq:SI g_M acts on U_int}
    \hat{g}_M^{int}\hat{U}_{int}(t,0)[\hat{g}_M^{int}]^{-1}&=\hat{U}_{int}(t+T/2)\hat{U}^{-1}(T/2)\hat{U}_0(T/2), \nonumber \\
    &=\hat{U}_{int}(t+T/2,T/2), 
\end{align}
and $[\hat{g}^{int}_M]^2=\hat{U}^{-1}_0(T)\hat{g}^2_M=\hat{X}^{-1}\hat{g}^2_M$. Therefore, $\hat{g}^{int}_M$ acts on $\hat{H}_{int}(t)$ as
\begin{equation} \label{eq:SI g_M subgroup of G_int}
    \hat{g}_M^{int}\hat{H}_{int}(t)[\hat{g}_M^{int}]^{-1}=\hat{H}_{int}(T/2+t).
\end{equation}

For $\hat{g}_A \in A\subset G_{st}$, defined as in Eq.~\eqref{eq:SI Definition of symmetries}, the symmetry transformations satisfy $\hat{g}_A\hat{U}(t,0)\hat{g}^{-1}_A=\hat{U}(T/2-t,T/2)$ and $\hat{g}_A\hat{U}_0(t,0)\hat{g}^{-1}_A=\hat{U}_0(T/2-t,T/2)$, which yield
\begin{align} \label{eq:SI g_A relations}
    \hat{g}_A\hat{U}_{int}(t)\hat{g}^{-1}_A&=\hat{U}_0(T/2,T/2-t)\hat{U}(T/2-t,T/2), \\
    &=\hat{U}_0(T/2)\hat{U}_{int}(T/2-t)\hat{U}^{-1}(T/2). \nonumber
\end{align}
Define the antiunitary symmetry element $\hat{g}^{int}_A\in G^{int}_{st}$ as $\hat{g}^{int}_A=\hat{U}^{-1}_0(T/2)\hat{g}_A$, which transforms $\hat{U}_{int}(t)$ as  
\begin{align} \label{eq:SI g_A acts on U_int}
    \hat{g}^{int}_A\hat{U}_{int}(t)[\hat{g}^{int}_A]^{-1}&=\hat{U}_{int}(T/2-t)\hat{U}^{-1}(T/2)\hat{U}_0(T/2), \nonumber \\
    &=\hat{U}_{int}(T/2-t,T/2),
\end{align}
and satisfies $[\hat{g}^{int}_A]^2=\hat{g}^2_A$. The action of $\hat{g}^{int}_A$ on $\hat{H}_{int}(t)$ follows 
\begin{equation} \label{eq:SI g_A subgroup of G_int}
    \hat{g}_A^{int}\hat{H}_{int}(t)[\hat{g}_A^{int}]^{-1}=\hat{H}_{int}(T/2-t).
\end{equation}
The algebraic structure of $G^{int}_{st}$ is further defined by 
\begin{equation} \label{eq:SI Commute between g_M and X and g_A and X}
    \hat{g}^{int}_M \hat{X} [\hat{g}^{int}_M]^{-1}=\hat{X},\quad \hat{g}^{int}_A \hat{X} [\hat{g}^{int}_A]^{-1}=\hat{X}^{-1},
\end{equation}
from 
\begin{align} \label{eq:SI group algebra of unitary G_int}
    \hat{g}_M^{int}\hat{X}[\hat{g}_M^{int}]^{-1}&=\hat{U}_0^{-1}(T/2) \hat{g}_M \hat{U}_0(T) [\hat{g}_M]^{-1}\hat{U}_0(T/2), \nonumber \\
    &=\hat{U}^{-1}_0(T/2)\hat{U}_0(T+T/2,T/2)\hat{U}_0(T/2), \nonumber \\ 
    &=\hat{X} 
\end{align}
and 
\begin{align} \label{eq:SI group algebra of antiunitary G_int}
    \hat{g}^{int}_A\hat{X}[\hat{g}^{int}_A]^{-1}&=\hat{U}^{-1}_0(T/2)\hat{g}_A\hat{X}\hat{g}^{-1}_A\hat{U}_0(T/2), \nonumber \\
    &=\hat{U}^{-1}_0(T/2)\hat{U}_0(-T/2,T/2)\hat{U}_0(T/2), \nonumber \\
    &=\hat{U}_0(0,T)=\hat{X}^{-1}. 
\end{align}
These relations establish the extended dynamical symmetry group in the interaction picture as $G^{int}_{st}=\mathbb{Z}_2 \times (\mathbb{Z}_N \rtimes \mathbb{Z}^{T}_2)$, encapsulating the interplay between the emergent $\mathbb{Z}_N$ symmetry and the original symmetry group $G_{st}=\mathbb{Z}_2\times\mathbb{Z}^{T}_2$.

\subsection{Symmetry group of the prethermal Hamiltonian}
The symmetry group of the truncated prethermal Hamiltonian $\hat{D}_n$ inherits symmetry elements $\hat{X}$, $\hat{g}^{int}_M$, and $\hat{g}^{int}_A$ from $G^{int}_{st}$ of $\hat{H}_{int}(t)$, as specified in Eqs.~\eqref{eq:SI X_Subgroup of G_int}, \eqref{eq:SI g_M subgroup of G_int}, and \eqref{eq:SI g_A subgroup of G_int}. The actions of these elements on the Fourier component $\hat{V}_m$ in Eq.~\eqref{eq: SI V_m} are expressed as
\begin{align} \label{eq:SI symmetry actions onto V_m}
    \hat{X}\hat{V}_m\hat{X}^{-1}&=e^{i2\pi m/N}\hat{V}_m, \\
    \hat{g}^{int}_{M/A}\hat{V}_m[\hat{g}^{int}_{M/A}]^{-1}&=e^{-i2\pi m/2N}\hat{V}_m. \nonumber 
\end{align}

The $i$-th order van Vleck expansion term, $\hat{V}^{[i]}_{vV}$, in $\hat{D}_n=\hat{V}^{vV,n}_{int}$ includes terms with numerators of the form $\hat{V}_{m_1}\hat{V}_{m_2}\cdots \hat{V}_{m_i}$, as shown in Eq.~\eqref{eq:SI vV expansions}. Symmetry operations act on these terms as follows
\begin{align} \label{eq:SI Symmetry actions onto van Vleck expansion of D_n}
    &\hat{X}\hat{V}_{m_1}\cdots \hat{V}_{m_i}\hat{X}^{-1}=e^{i\frac{2\pi}{N}(\sum^{i}_{k=1}m_k)}\hat{V}_{m_1}\cdots \hat{V}_{m_i}, \\
    &\hat{g}^{int}_{M/A}\hat{V}_{m_1}\cdots \hat{V}_{m_i}[\hat{g}^{int}_{M/A}]^{-1}=e^{-i\frac{2\pi}{2N}(\sum^{i}_{k=1}m_k)}\hat{V}_{m_1}\cdots \hat{V}_{m_i}. \nonumber 
\end{align}
The van Vleck gauge condition imposes a constraint on the indices, $\sum^{i}_{k=1}m_k=0$~\cite{Mikami2016,Mizuta2019}, leading to the commutation relations
\begin{equation} \label{eq:SI Symmetries of prethermal Hamiltonian D_n}
    [\hat{X}, \hat{D}_n]=[\hat{g}^{int}_M, \hat{D}_n]=[\hat{g}^{int}_A, \hat{D}_n]=0,
\end{equation}
valid for any integer $n\leq n_{0}=O(\tilde{\Omega})$. Thus, the prethermal Hamiltonian $\hat{D}_n$ inherits the full symmetry group $G^{int}_{st}$ of $\hat{H}_{int}(t)$.

From Eqs.~\eqref{eq:SI vV expansions} and \eqref{eq:SI symmetry actions onto V_m}, the transformations of $\hat{K}^{[i]}(t)$ under symmetry operations are given by 
\begin{align} \label{eq:SI Symmetries actions onto Kick}
    \hat{X}\hat{K}^{[i]}(t)\hat{X}^{-1}&=\hat{K}^{[i]}(t-mT), \\
    \hat{g}^{int}_{M/A}\hat{K}^{[i]}(t)[\hat{g}^{int}_{M/A}]^{-1}&=\pm\hat{K}^{[i]}(T/2\pm t). \nonumber
\end{align}
These relations yield the following transformations presented in the main text, $\hat{X}^m\mathcal{\hat{U}}_n(t)\hat{X}^{-m}=\mathcal{\hat{U}}_n(t-mT)$ and $\hat{g}^{int}_{M/A}\mathcal{\hat{U}}_n(t)[\hat{g}^{int}_{M/A}]^{-1}=\mathcal{\hat{U}}_n(T/2\pm t)$. The time evolution over $lT$ periods, with $l\bmod N$, is expressed as $\hat{U}(lT)=\hat{X}^{l}\hat{U}_{int}\simeq \hat{X}^{l}\mathcal{\hat{U}}_n(lT)e^{-i\hat{D}_n lT}\mathcal{\hat{U}}^{\dag}_n(0)=\mathcal{\hat{U}}_n(0)\hat{X}^le^{-i\hat{D}_n lT}\mathcal{\hat{U}}^{\dag}_n(0)$, which matches the expression in Ref.~\cite{Else2017_1}. In addition, the symmetry transformations for $\hat{U}_{int}(t_2,t_1)$ follow as
\begin{align}
    &\hat{X}\hat{U}_{int}(t_2,t_1)\hat{X}^{-1}=\hat{X}[\mathcal{\hat{U}}_n(t_2)e^{-i\hat{D}_n (t_2-t_1)}\mathcal{\hat{U}}^{\dag}_n(t_1)]\hat{X}^{-1} \nonumber \\
    &=\mathcal{\hat{U}}_n(t_2-T)\hat{X}e^{-i\hat{D}_n(t_2-t_1)}\hat{X}^{-1}\mathcal{\hat{U}}^{\dag}_n(t_1-T) \nonumber \\
    &=\hat{U}_{int}(t_2-T,t_1-T),
\end{align}
and 
\begin{align}
    &\hat{g}^{int}_{M/A}\hat{U}_{int}(t_2,t_1)[\hat{g}^{int}_{M/A}]^{-1} \nonumber \\
    &=\mathcal{\hat{U}}_n(T/2\pm t_2)\hat{g}^{int}_{M/A}e^{-i\hat{D}_n (t_2-t_1)}[\hat{g}^{int}_{M/A}]^{-1}\mathcal{\hat{U}}^{\dag}_{n}(T/2\pm t_1) \nonumber \\
    &=\hat{U}_{int}(T/2\pm t_2,T/2\pm t_1),
\end{align}
using $[\hat{X},\hat{D}_n]=[\hat{g}^{int}_{M/A},\hat{D}_n]=0$. These results are equivalent to Eqs.~\eqref{eq:SI X_Subgroup of G_int}, \eqref{eq:SI g_M acts on U_int} and \eqref{eq:SI g_A acts on U_int}. 

\section{Detection of dynamical symmetry}
This section introduces a scheme to engineer nontrivial micromotion during the prethermal stage, using dynamical symmetries in $G_{st}$. Specifically, we focus on the order-$2$ unitary symmetry $\hat{g}_M\in G_{st}$ of $\hat{H}(t)$ in Eq.~\eqref{eq:SI Definition of symmetries}, which can be generalized to higher-order symmetries. 

\subsection{Derivation of key symmetry relations}
We derive key relations for $\hat{g}^{\prime}_M\mathcal{\hat{U}}_n(0)[\hat{g}^{int}_M]^{-1}$ and $\hat{g}^{\prime -1}_M\mathcal{\hat{U}}_n(0)\hat{g}^{int}_M$, which are essential for developing schemes to detect dynamical symmetries. Here, $\hat{g}^{\prime}_M=\hat{U}^{-1}(T/2)\hat{g}_M$ is a projective symmetry element of the extended symmetry group for the Floquet operator $\hat{U}(T)$, satisfying $\hat{g}^{\prime 2}_M=\hat{U}^{-1}(T)\hat{g}^2_M$, where $\hat{g}^2_M=e^{i\beta}\mathbb{I}$, as introduced in Ref.~\cite{Na2023}. The expression for $\hat{g}^{\prime}_M\mathcal{\hat{U}}_n(0)[\hat{g}^{int}_M]^{-1}$ is obtained as follows 
\begin{align}  \label{eq:SI Expression 1 (a)}
    \hat{g}^{\prime}_M\mathcal{\hat{U}}_n(0)[\hat{g}^{int}_M]^{-1}&=\hat{g}_M[\hat{g}^{int}_M]^{-1}\mathcal{\hat{U}}_n(T/2) \\
    &=\hat{U}^{-1}(T/2)\hat{U}_0(T/2)\mathcal{\hat{U}}_n(T/2) \nonumber \\
    &=\hat{U}_{int}(0,T/2)\mathcal{\hat{U}}_n(T/2)=\mathcal{\hat{U}}_n(0)e^{i\hat{D}_nT/2}. \nonumber
\end{align}
Now, consider the expression $\hat{g}^{\prime -1}_M\mathcal{\hat{U}}_n(0)\hat{g}^{int}_M$. It is derived as
\begin{align} \label{eq:SI Expression 1 (b)}
    &[\hat{g}^{int}_M\mathcal{\hat{U}}^{\dag}_n(0)\hat{g}^{\prime -1}_M][\hat{g}^{\prime -1}_M\mathcal{\hat{U}}_n(0)\hat{g}^{int}_M]=\hat{g}^{int}_M\mathcal{\hat{U}}^{\dag}_n(0)\hat{g}^{\prime -2}_M\mathcal{\hat{U}}_n(0)\hat{g}^{int}_M \nonumber \\
    &=\hat{g}^{int}_M\mathcal{\hat{U}}^{\dag}_n(0)\hat{g}^{-2}_M\hat{U}(T)\mathcal{\hat{U}}_n(0)\hat{g}^{int}_M=e^{-i\beta}\hat{g}^{int}_M\hat{X}e^{-i\hat{D}_nT}\hat{g}^{int}_M \nonumber \\
    &=e^{-i\beta}\hat{X}e^{-i\hat{D}_nT}[\hat{g}^{int}_M]^2=e^{-i\hat{D}_nT}, 
\end{align}
using $[\hat{g}^{int}_M,\hat{X}]=[\hat{g}^{int}_M,\hat{D}_n]=[\hat{X},\hat{D}_n]=0$ and $[\hat{g}^{int}_M]^2=\hat{X}^{-1}\hat{g}^2_M=e^{i\beta}\hat{X}^{-1}$. 
Combining this result with Eq.~\eqref{eq:SI Expression 1 (a)}, we obtain the expression $\hat{g}^{\prime -1}_M\mathcal{\hat{U}}_n(0)\hat{g}^{int}_M$ as
\begin{align} \label{eq:SI Expression 1 (c)}
    \hat{g}^{\prime -1}_M\mathcal{\hat{U}}_n(0)\hat{g}^{int}_M&=[\hat{g}^{int}_M\mathcal{\hat{U}}^{\dag}_n(0)\hat{g}^{\prime -1}_M]^{-1}e^{-i\hat{D}_n T}\\
    &=[\mathcal{\hat{U}}_n(0)e^{i\hat{D}_n T/2}]e^{-i\hat{D}_nT}=\mathcal{\hat{U}}_n(0)e^{-i\hat{D}_n T/2}.  \nonumber
\end{align}

\subsection{Detection scheme for the initial state: $[\hat{\rho}(0),\hat{D}_n]=0$}
For a set of Hermitian operators $\hat{O}$ that satisfy the relation $\hat{g}_M\hat{O}\hat{g}^{-1}_M=e^{i\alpha_M}\hat{O}$, where $e^{i\alpha_M}=\pm1$ due to Hermiticity, we examine the expectations values $\langle\hat{O}(mT)\rangle$ and %=\Tr[\hat{\rho}(mT)\hat{O}_i]$ 
$\langle\hat{O}(mT+T/2)\rangle$ for integer $m$ in the prethermal regime, and derive the relation between them imposed by the presence of dynamical symmetries. The expectation value at $t=mT+T/2$ is given by
\begin{align} \label{eq:SI expectation values of O at mT+T/2 (a)}
    &\langle \hat{O}(mT+T/2)\rangle=\Tr[\hat{U}(T/2)\hat{\rho}(mT)\hat{U}^{-1}(T/2)\hat{O}] \nonumber \\
    &=e^{-i\alpha_M}\Tr[\hat{g}^{-1}_M\hat{U}(T/2)\hat{\rho}(mT)\hat{U}^{-1}(T/2)\hat{g}_M\hat{O}] \nonumber \\
    &=e^{-i\alpha_M}\Tr[\hat{g}^{\prime -1}_M\hat{\rho}(mT)\hat{g}^{\prime}_M\hat{O}],
\end{align}
where $\hat{\rho}(t)$ is the density matrix of the system at $t$. By applying $\hat{g}^{\prime -1}_M=\mathcal{\hat{U}}_n(0)e^{-i\hat{D}_n T/2}[\hat{g}^{int}_M]^{-1}\mathcal{\hat{U}}^{\dag}_n(0)$ derived in Eq.~\eqref{eq:SI Expression 1 (c)} and $\hat{U}(mT)\simeq \mathcal{\hat{U}}_n(0)\hat{X}^me^{-i\hat{D}_n mT}\mathcal{\hat{U}}^{\dag}_n(0)$, the term $\hat{g}^{\prime -1}_M\hat{\rho}(mT)\hat{g}^{\prime}_M$ simplifies to $\hat{A}\hat{\rho}(0)\hat{A}^{-1}$, where
\begin{equation} \label{eq:SI operator A}
    \hat{A}=\mathcal{\hat{U}}_n(0)e^{-i\hat{D}_n T/2}[\hat{g}^{int}_M]^{-1}\hat{X}^me^{-i\hat{D}_n mT}\mathcal{\hat{U}}^{\dag}_n(0).
\end{equation}
Approximating $\mathcal{\hat{U}}^{\dag}_n(0)\hat{\rho}(0)\mathcal{\hat{U}}_n(0)\simeq\hat{\rho}(0)+i[\hat{K}^{[1]}(0),\hat{\rho}(0)]+O(\tilde{\Omega}^{-2})\hat{\rho}(0)\simeq\hat{\rho}(0)$ and using $[\hat{g}^{int}_M,\hat{D}_n]=0$, we obtain
\begin{align} \label{eq:SI expectation values of O at mT+T/2 (b)}
    \langle \hat{O}(mT+T/2)\rangle &\simeq e^{-i\alpha_M}\Tr\left[ \hat{\rho}(mT)\hat{O}\right ] \\
    &=e^{-i\alpha_M}\langle \hat{O}(mT) \rangle, \nonumber
\end{align}
if the initial state $\ket{\Psi(0)}$ is taken as the eigenstate of the truncated prethermal Hamiltonian $\hat{D}_n$ for $n\leq n_0$. 
\begin{figure}
    \centering
    \includegraphics[width=\columnwidth]{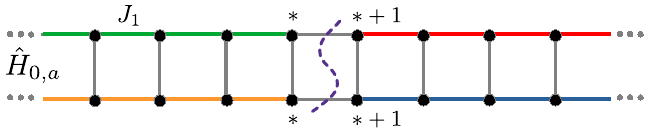}
    \caption{Schematic of one-dimensional spin-$1/2$ ladder of length $L$. $S$ spins reside in the upper chain and $\sigma$ spins in the lower chain. Along each chain, the spins interact via nearest neighbor couplings with strength $J_1\sim \Omega$, as described in $\hat{H}_{0,a}$ of Eq.~\eqref{eq:SI four-step driven model}. The symmetry operation $\hat{g}_M$ represents a mirror reflection about the centerline of the ladder. Here, the notation $*=L/2-1$ denotes the left sites of the upper and lower chains adjacent to the center bond of the ladder.}
    \label{fig:SI model}
\end{figure}

\subsection{Detection scheme for the initial state: $[\hat{\rho}(0),\hat{D}_n]\neq0$}
We generalize the above detection scheme to be independent of the initial state, making it highly amenable to experimental realization, by imposing additional conditions on the chosen Hermitian operator $\hat{O}$, which satisfies $[\hat{g}^{int}_M,\hat{O}]=[\hat{D}_0,\hat{O}]=0$. Starting from Eq.~\eqref{eq:SI operator A} and approximating $\mathcal{\hat{U}}_n(0)\sim\mathbb{I}$, we obtain $\langle \hat{O}(mT+T/2)\rangle \simeq e^{-i\alpha_M}\Tr[\hat{X}^m e^{-i\hat{D}_n mT}\hat{\rho}(0) e^{i\hat{D}_n mT}\hat{X}^{-m}\hat{B}]$, where 
\begin{equation} \label{eq:SI operator B}
    \hat{B}=\hat{g}^{int}_Me^{i\hat{D}_nT/2}\hat{O}e^{-i\hat{D}_nT/2}[\hat{g}^{int}_M]^{-1}.
\end{equation}
For comparison, the expectation value at $t=mT$ is $\langle \hat{O}(mT)\rangle\simeq \Tr[\hat{X}^m e^{-i\hat{D}_n mT}\hat{\rho}(0)e^{i\hat{D}_n mT}\hat{X}^{-m}\hat{O}]$. Using $[\hat{g}^{int}_M,\hat{O}]=0$ and the approximation $[\hat{D}_n,\hat{O}]\simeq0$ valid for $1\ll \tilde{\Omega}$, we find $\hat{B}\simeq \mathbb{I}$, leading to the following relation between the two expressions as $\langle \hat{O}(mT+T/2)\rangle \simeq e^{-i\alpha_M}\langle \hat{O}(mT)\rangle$, which is equivalent to Eq.~\eqref{eq:SI expectation values of O at mT+T/2 (b)}. 

\section{Additional details on numerics}
\begin{figure}
    \centering
    \includegraphics[width=\columnwidth]{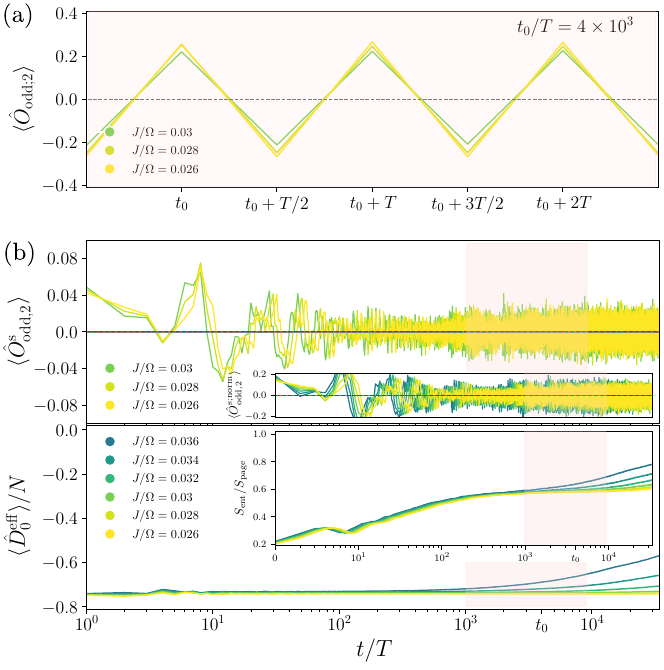}
    \caption{After relaxation, the system enters a prethermal regime governed approximately by the effective prethermal Hamiltonian $\hat{D}_0$ in Eq.~\eqref{eq:SI D_0} until $t_{0}=n_0T$ ($n_0=O(\tilde{\Omega})$), with $t_0$ marking a reference temporal point within this regime. (a) Time evolution of $\langle \hat{O}_{\rm odd,2}(t)\rangle$ at $t=mT$ and $t=mT+T/2$ within the prethermal regime verifies Eq.~\eqref{eq:SI expectation values of O at mT+T/2 (b)} and detects $\hat{g}_M$. The shaded regions indicate the prethermal regime, with bipartite entanglement and energy density consistent across a range of frequencies ($J/\Omega=0.026,0.028$). (b) Time evolution of $\hat{O}^{s}_{\rm odd,2}(mT)=\hat{O}_{\rm odd,2}(mT)+\hat{O}_{\rm odd,2}(mT+T/2)$ stabilizes at zero (red horizontal line) in the prethermal regime, confirming the presence of $\hat{g}_M$. The normalized quantity $\hat{O}^{s;\rm norm}_{\rm odd,2}(mT)=\hat{O}^{s}_{\rm odd,2}(mT)/||\hat{O}_{\rm odd,2}(mT)||$, shown in the inset, distinguishes symmetry-preserving prethermal dynamics from thermalizing states approaching infinite temperature. The parameters used match Fig. 1 in the main text.}
    \label{fig:SI Prethermal dynamics}
\end{figure}
\begin{figure*}
    \centering
    \includegraphics[width=\textwidth]{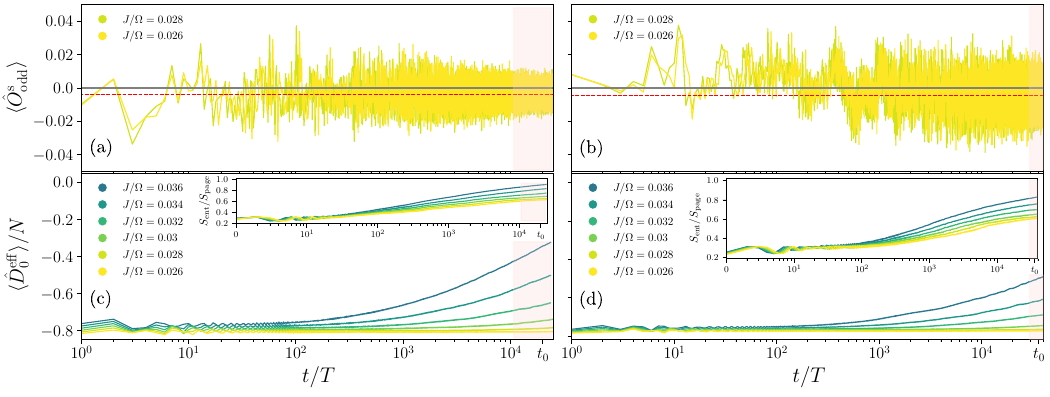}
    \caption{In the symmetry-broken cases (a) $(\lambda^{LR}_a,\lambda^{LR}_b)=(0.8,1.2)$ and (b) $(\lambda^{LR}_a,\lambda^{LR}_b)=(0.5,1)$, $\hat{O}^{s}_{\rm odd}(mT)$ stabilizes at a non-zero value, signaling symmetry breaking. The red horizontal line denotes the center of stable oscillations within the prethermal regime. With increased symmetry breaking,  the offset from zero becomes more pronounced. (c) and (d) Energy density and bipartite entanglement entropy (insets) indicate the prethermal regime, highlighted by shaded regions. The parameters used are consistent with Fig. 1, except for the symmetry-breaking values of $(\lambda^{LR}_a, \lambda^{LR}_b)$. }
    \label{fig:SI Symmetry broken cases}
\end{figure*}
\subsection{Model}
As a concrete example, we introduce a model of a two-leg, spin-$1/2$ ladder of length $L$ under dual driving, with an order-2 symmetry, $\hat{g}_M\in G_{st}=\mathbb{Z}_2$. Specifically, $\hat{g}_M$ represents a mirror operation along the centerline of the ladder, as shown in Fig.~\ref{fig:SI model}. The spins in the upper (lower) chain are labeled by $S~(\sigma)$. We design a four-step driven model defined by 
\begin{equation} \label{eq:SI four-step driven model}
    \hat{H}(t)=\begin{cases}
         \hat{H}_{0,a}&(0 \leq t < \tau T), \\ 
         \hat{V}_{a}+\hat{V}^{sc}_{a}&(\tau T \leq t < T/2), \\ 
         \hat{H}_{0,b}&(T/2 \leq t < \tau T+T/2),\\ 
         \hat{V}_{b}+\hat{V}^{sc}_{b}&(\tau T+T/2 \leq t < T), 
         \end{cases}
\end{equation}
satisfying the symmetry relations
$\hat{g}_M\hat{H}_{0,a}\hat{g}^{-1}_M=\hat{H}_{0,b}$ and $\hat{g}_M\hat{V}_{a}\hat{g}^{-1}_M=\hat{V}_{b}$. The resonant $\hat{H}_{0,a/b}$ and the weak drive $\hat{V}_{a/b}$ are provided in the main text. The coupling strength $J_1=\Omega/4\tau$ ensures that the resonant drive $\hat{H}_{0}(t)$ generates the $\mathbb{Z}_2$ symmetry operation $\hat{X}=e^{-i\hat{H}_{0,b}\tau T}e^{-i\hat{H}_{0,a}\tau T}$, where $\hat{X}^2=\mathbb{I}$. For consistency with prior works~\cite{White2018,Rakovszky2022} on hydrodynamics, we adopt the following parameter values for the high-frequency drive $\hat{V}_{a/b}$: $g_x=g_y=0.45225$, and $g_z=0.7$. We set $\tau=0.25$ and $J=1$, leading to $J^{\prime}=J/(0.5-\tau)=4$. 

We introduce an additional coupling $\hat{V}_{a/b}^{sc}$ at the center of the ladder
\begin{align} \label{eq:SI V sym breaking}
&\hat{V}^{sc}_{a/b}=J^{\prime}\lambda^{LR}_{a/b}[S^{x}_{*}S^{x}_{*+1}+\sigma^y_{*}\sigma^y_{*+1}+g_{zz}S^z_{*+1/*}\sigma^z_{*+1/*} \nonumber \\
&+g_{x}S^x_{*+1/*}+g_{z}S^z_{*+1/*}+g_{y}\sigma^y_{*+1/*}+g_{z}\sigma^z_{*+1/*}], 
\end{align}
where $*=L/2-1$ (see Fig.~\ref{fig:SI model}). This allows symmetry breaking by setting $\lambda^{LR}_{a}\neq \lambda^{LR}_{b}$, causing $\hat{g}_M\hat{V}^{sc}_a\hat{g}^{-1}_M\neq\hat{V}^{sc}_b$. For the numerical results in the main text, we set $(\lambda^{LR}_{a},\lambda^{LR}_b)=(0.5,0.5)$ for Fig. 1 and $(1,1)$ and $(0.8,1.2)$ in Fig. 2(a) and (b). Additionally, we set an Ising interaction strength $g_{zz}=1.3$ to couple the spins across each rung of the ladder. 

For the model described in Eq.~\eqref{eq:SI four-step driven model}, the symmetry operator $\hat{g}_M$ is given by
\begin{equation} \label{eq:SI Symmetry g_M}
    \hat{g}_M=\prod^{L/2-1}_{i=0}e^{-i\frac{\pi}{4}(\mathbb{I}+\vec{S}_{i}\cdot\vec{S}_{L-1-i})}e^{-i\frac{\pi}{4}(\mathbb{I}+\vec{\sigma}_{i}\cdot\vec{\sigma}_{L-1-i})},
\end{equation}
and the symmetry operator $\hat{X}$ takes the form 
\begin{equation} \label{eq:SI symmetry X model}
    \hat{X}=[S^{x}_0S^{x}_{*}S^{x}_{*+1}S^{x}_{L-1}][\sigma^{y}_0\sigma^{y}_{*}\sigma^{y}_{*+1}\sigma^{y}_{L-1}],\quad\hat{X}^2=\mathbb{I}.
\end{equation}
The zeroth-order effective prethermal Hamiltonian, $\hat{D}_0=[\int^{T}_0dt\hat{V}_{int}(t)+\hat{X}^{-1}\int^{T}_0dt\hat{V}_{int}(t)\hat{X}]/2T$, can be decomposed as $\hat{D}_0=\hat{D}^{L}_0+\hat{D}^{R}_0+\hat{D}^{c}_0$, where $\hat{D}^{L}_0$, $\hat{D}^{R}_0$, and $\hat{D}^{c}_0$ correspond to the left half, right half, and center of the ladder, respectively, as depicted in Fig.~\ref{fig:SI model}. These terms are defined as follows
\begin{align} 
    &\hat{D}^{L}_0=J[-S^x_0S^x_1-\sigma^y_0\sigma^y_1+\sum^{L/2-3}_{i=1}(S^x_iS^x_{i+1}+\sigma^y_i\sigma^y_{i+1}) \nonumber \\
    &-S^x_{*-1}S^x_{*}-\sigma^y_{*-1}\sigma^y_{*}-g_{zz}S^z_0\sigma^z_0+\sum^{L/2-2}_{i=1}g_{zz}S^z_i\sigma^z_i \nonumber \\
    &-g_{x}S^{x}_0-g_{y}\sigma^{y}_0+\sum^{L/2-2}_{i=1}(g_{x}S^x_i+g_{z}S^z_i+g_{y}\sigma^y_i+g_{z}\sigma^z_i)], \nonumber 
\end{align}
\begin{align} 
    &\hat{D}^{R}_0=J[\sum^{L-2}_{i=L/2}(S^x_iS^x_{i+1}+\sigma^y_i\sigma^y_{i+1})+\sum^{L-1}_{i=L/2+1}g_{zz}S^z_i\sigma^z_i+ \nonumber \\
    &\sum^{L-2}_{i=L/2+1}(g_{x}S^x_i+g_{z}S^z_i+g_{y}\sigma^y_i+g_{z}\sigma^z_i)+g_xS^{x}_{L-1}+g_y\sigma^{y}_{L-1}], \nonumber 
\end{align}
and
\begin{align} \label{eq:SI D_0}
    &\hat{D}^{c}_0=J[(\lambda^{LR}_a+\lambda^{LR}_b)+(S^x_{*}S^x_{*+1}+\sigma^y_{*}\sigma^y_{*+1})+ \\
    &(\lambda^{LR}_b-1)g_{zz}S^z_{*}\sigma^z_{*}+(1-\lambda^{LR}_a)g_{zz}S^{z}_{*+1}\sigma^z_{*+1}+ \nonumber \\
    &(\lambda^{LR}_b-1)(g_xS^{x}_{*}+g_{y}\sigma^{y}_{*})+(1-\lambda^{LR}_a)(g_xS^{x}_{*+1}+g_y\sigma^{y}_{*+1})]. \nonumber
\end{align}

\subsection{Dynamical symmetry-preserving prethermalization}
In the main text, to probe the dynamical symmetry, we define the odd operator $\hat{O}_{\rm odd}=S^{x}_{L/2-1}-S^{x}_{L/2}$, which satisfies $\hat{g}_M\hat{O}_{\rm odd}\hat{g}^{-1}_M=-\hat{O}_{\rm odd}$ with $e^{i\alpha_M}=-1$. We also impose the conditions $[\hat{g}^{int}_M,\hat{O}_{\rm odd}]=[\hat{D}_0,\hat{O}_{\rm odd}]=0$, ensuring that the detection scheme is independent of the initial state: $[\hat{\rho}(0),\hat{D}_0]\neq0$. The expected relation $\langle \hat{O}_{\rm odd}(mT+T/2)\rangle=-\langle \hat{O}_{\rm odd}(mT)\rangle$ in Eq.~\eqref{eq:SI expectation values of O at mT+T/2 (b)} is verified through the time evolution of these operators, demonstrating the emergence of nontrivial micromotion induced by the dynamical symmetry $\hat{g}_M$ within the prethermal regime (see Fig. 1 in the main text).  

Here, we confirm this result for another odd operator, $\hat{O}_{\rm odd,2}=\sigma^{y}_{L/2-1}-\sigma^{y}_{L/2}$ [Fig.~\ref{fig:SI Prethermal dynamics}(a)]. We define $\hat{O}^{s}_{\rm odd,2}(mT)=\hat{O}_{\rm odd,2}(mT)+\hat{O}_{\rm odd,2}(mT+T/2)$ for integer $m$, and observe that $\hat{O}^{s}_{\rm odd,2}(t)$ saturates to zero, as indicated by the red horizontal line as the system enters the prethermal regime following relaxation [Fig.~\ref{fig:SI Prethermal dynamics}(b)]. The shaded region marks the prethermal regime for $J/\Omega=0.026, 0.028$, while for $J/\Omega=0.03$, the dynamics exit the prethermal regime. Here, $t_0$ serves as a reference temporal point within this regime. 

\subsection{Sensitivity to dynamical symmetry breaking}
To verify the perfect oscillation of $\langle\hat{O}_{\rm odd}(mT)\rangle = -\langle \hat{O}_{\rm odd}(mT+T/2)\rangle$ is directly tied to the dynamical symmetry, we explicitly break the symmetry $\hat{g}_M$, by adjusting the coupling across the center bond by setting $\lambda^{LR}_a\neq\lambda^{LR}_b$ in Eq.~\eqref{eq:SI V sym breaking}. When the symmetry is broken by setting $(\lambda^{LR}_a,\lambda^{LR}_b)=(0.8,1.2)$, the saturated value of $\hat{O}^{s}_{\rm odd}(t)$ is nonzero, as shown by the location of the red horizontal line in Fig.~\ref{fig:SI Symmetry broken cases}(a). As the symmetry breaking increases to $(\lambda^{LR}_a,\lambda^{LR}_b)=(0.5,1)$, the offset from zero becomes larger, as seen in Fig.~\ref{fig:SI Symmetry broken cases}(b).
%\end{widetext}
\bibliography{manuscript}  % <-- keep your existing .bib name

@article{Kitagawa2010,
  title = {Topological characterization of periodically driven quantum systems},
  author = {Kitagawa, Takuya and Berg, Erez and Rudner, Mark and Demler, Eugene},
  journal = {Phys. Rev. B},
  volume = {82},
  issue = {23},
  pages = {235114},
  numpages = {12},
  year = {2010},
  month = {Dec},
  publisher = {American Physical Society},
  doi = {10.1103/PhysRevB.82.235114},
  url = {https://link.aps.org/doi/10.1103/PhysRevB.82.235114}
}

@article{Rudner2013,
  title = {Anomalous Edge States and the Bulk-Edge Correspondence for Periodically Driven Two-Dimensional Systems},
  author = {Rudner, Mark S. and Lindner, Netanel H. and Berg, Erez and Levin, Michael},
  journal = {Phys. Rev. X},
  volume = {3},
  issue = {3},
  pages = {031005},
  numpages = {15},
  year = {2013},
  month = {Jul},
  publisher = {American Physical Society},
  doi = {10.1103/PhysRevX.3.031005},
  url = {https://link.aps.org/doi/10.1103/PhysRevX.3.031005}
}

@article{Else2016_1,
  title = {Classification of topological phases in periodically driven interacting systems},
  author = {Else, Dominic V. and Nayak, Chetan},
  journal = {Phys. Rev. B},
  volume = {93},
  issue = {20},
  pages = {201103},
  numpages = {5},
  year = {2016},
  month = {May},
  publisher = {American Physical Society},
  doi = {10.1103/PhysRevB.93.201103},
  url = {https://link.aps.org/doi/10.1103/PhysRevB.93.201103}
}

@article{Titum2016,
  title = {Anomalous Floquet-Anderson Insulator as a Nonadiabatic Quantized Charge Pump},
  author = {Titum, Paraj and Berg, Erez and Rudner, Mark S. and Refael, Gil and Lindner, Netanel H.},
  journal = {Phys. Rev. X},
  volume = {6},
  issue = {2},
  pages = {021013},
  numpages = {20},
  year = {2016},
  month = {May},
  publisher = {American Physical Society},
  doi = {10.1103/PhysRevX.6.021013},
  url = {https://link.aps.org/doi/10.1103/PhysRevX.6.021013}
}

@article{Keyserlingk2016_1,
  title = {Phase structure of one-dimensional interacting Floquet systems. I. Abelian symmetry-protected topological phases},
  author = {von Keyserlingk, C. W. and Sondhi, S. L.},
  journal = {Phys. Rev. B},
  volume = {93},
  issue = {24},
  pages = {245145},
  numpages = {18},
  year = {2016},
  month = {Jun},
  publisher = {American Physical Society},
  doi = {10.1103/PhysRevB.93.245145},
  url = {https://link.aps.org/doi/10.1103/PhysRevB.93.245145}
}

@article{Keyserlingk2016_2,
  title = {Phase structure of one-dimensional interacting Floquet systems. II. Symmetry-broken phases},
  author = {von Keyserlingk, C. W. and Sondhi, S. L.},
  journal = {Phys. Rev. B},
  volume = {93},
  issue = {24},
  pages = {245146},
  numpages = {11},
  year = {2016},
  month = {Jun},
  publisher = {American Physical Society},
  doi = {10.1103/PhysRevB.93.245146},
  url = {https://link.aps.org/doi/10.1103/PhysRevB.93.245146}
}

@article{Potter2016_1,
  title = {Classification of Interacting Topological Floquet Phases in One Dimension},
  author = {Potter, Andrew C. and Morimoto, Takahiro and Vishwanath, Ashvin},
  journal = {Phys. Rev. X},
  volume = {6},
  issue = {4},
  pages = {041001},
  numpages = {19},
  year = {2016},
  month = {Oct},
  publisher = {American Physical Society},
  doi = {10.1103/PhysRevX.6.041001},
  url = {https://link.aps.org/doi/10.1103/PhysRevX.6.041001}
}

@article{Harper2017,
  title = {Floquet Topological Order in Interacting Systems of Bosons and Fermions},
  author = {Harper, Fenner and Roy, Rahul},
  journal = {Phys. Rev. Lett.},
  volume = {118},
  issue = {11},
  pages = {115301},
  numpages = {6},
  year = {2017},
  month = {Mar},
  publisher = {American Physical Society},
  doi = {10.1103/PhysRevLett.118.115301},
  url = {https://link.aps.org/doi/10.1103/PhysRevLett.118.115301}
}

@article{Po2017,
  title = {Radical chiral Floquet phases in a periodically driven Kitaev model and beyond},
  author = {Po, Hoi Chun and Fidkowski, Lukasz and Vishwanath, Ashvin and Potter, Andrew C.},
  journal = {Phys. Rev. B},
  volume = {96},
  issue = {24},
  pages = {245116},
  numpages = {10},
  year = {2017},
  month = {Dec},
  publisher = {American Physical Society},
  doi = {10.1103/PhysRevB.96.245116},
  url = {https://link.aps.org/doi/10.1103/PhysRevB.96.245116}
}

@article{Fidkowski2019,
  title = {Interacting invariants for Floquet phases of fermions in two dimensions},
  author = {Fidkowski, Lukasz and Po, Hoi Chun and Potter, Andrew C. and Vishwanath, Ashvin},
  journal = {Phys. Rev. B},
  volume = {99},
  issue = {8},
  pages = {085115},
  numpages = {18},
  year = {2019},
  month = {Feb},
  publisher = {American Physical Society},
  doi = {10.1103/PhysRevB.99.085115},
  url = {https://link.aps.org/doi/10.1103/PhysRevB.99.085115}
}

@article{Khemani2016,
  title = {Phase Structure of Driven Quantum Systems},
  author = {Khemani, Vedika and Lazarides, Achilleas and Moessner, Roderich and Sondhi, S. L.},
  journal = {Phys. Rev. Lett.},
  volume = {116},
  issue = {25},
  pages = {250401},
  numpages = {6},
  year = {2016},
  month = {Jun},
  publisher = {American Physical Society},
  doi = {10.1103/PhysRevLett.116.250401},
  url = {https://link.aps.org/doi/10.1103/PhysRevLett.116.250401}
}

@article{Else2016_2,
  title = {Floquet Time Crystals},
  author = {Else, Dominic V. and Bauer, Bela and Nayak, Chetan},
  journal = {Phys. Rev. Lett.},
  volume = {117},
  issue = {9},
  pages = {090402},
  numpages = {5},
  year = {2016},
  month = {Aug},
  publisher = {American Physical Society},
  doi = {10.1103/PhysRevLett.117.090402},
  url = {https://link.aps.org/doi/10.1103/PhysRevLett.117.090402}
}

@article{Yao2017,
  title = {Discrete Time Crystals: Rigidity, Criticality, and Realizations},
  author = {Yao, N. Y. and Potter, A. C. and Potirniche, I.-D. and Vishwanath, A.},
  journal = {Phys. Rev. Lett.},
  volume = {118},
  issue = {3},
  pages = {030401},
  numpages = {6},
  year = {2017},
  month = {Jan},
  publisher = {American Physical Society},
  doi = {10.1103/PhysRevLett.118.030401},
  url = {https://link.aps.org/doi/10.1103/PhysRevLett.118.030401}
}

@article{Else2020,
   author = "Else, Dominic V. and Monroe, Christopher and Nayak, Chetan and Yao, Norman Y.",
   title = "Discrete Time Crystals", 
   journal= "Annual Review of Condensed Matter Physics",
   year = "2020",
   volume = "11",
   number = "Volume 11, 2020",
   pages = "467-499",
   doi = "https://doi.org/10.1146/annurev-conmatphys-031119-050658",
   url = "https://www.annualreviews.org/content/journals/10.1146/annurev-conmatphys-031119-050658",
   publisher = "Annual Reviews",
   issn = "1947-5462",
   type = "Journal Article",
}

@article{Zaletel2023,
  title = {Colloquium: Quantum and classical discrete time crystals},
  author = {Zaletel, Michael P. and Lukin, Mikhail and Monroe, Christopher and Nayak, Chetan and Wilczek, Frank and Yao, Norman Y.},
  journal = {Rev. Mod. Phys.},
  volume = {95},
  issue = {3},
  pages = {031001},
  numpages = {34},
  year = {2023},
  month = {Jul},
  publisher = {American Physical Society},
  doi = {10.1103/RevModPhys.95.031001},
  url = {https://link.aps.org/doi/10.1103/RevModPhys.95.031001}
}

@article{Mizuta2020,
  title = {Exact Floquet quantum many-body scars under Rydberg blockade},
  author = {Mizuta, Kaoru and Takasan, Kazuaki and Kawakami, Norio},
  journal = {Phys. Rev. Res.},
  volume = {2},
  issue = {3},
  pages = {033284},
  numpages = {13},
  year = {2020},
  month = {Aug},
  publisher = {American Physical Society},
  doi = {10.1103/PhysRevResearch.2.033284},
  url = {https://link.aps.org/doi/10.1103/PhysRevResearch.2.033284}
}

@article{Sugiura2021,
  title = {Many-body scar state intrinsic to periodically driven system},
  author = {Sugiura, Sho and Kuwahara, Tomotaka and Saito, Keiji},
  journal = {Phys. Rev. Res.},
  volume = {3},
  issue = {1},
  pages = {L012010},
  numpages = {6},
  year = {2021},
  month = {Feb},
  publisher = {American Physical Society},
  doi = {10.1103/PhysRevResearch.3.L012010},
  url = {https://link.aps.org/doi/10.1103/PhysRevResearch.3.L012010}
}

@article{Maskara2021,
  title = {Discrete Time-Crystalline Order Enabled by Quantum Many-Body Scars: Entanglement Steering via Periodic Driving},
  author = {Maskara, N. and Michailidis, A. A. and Ho, W. W. and Bluvstein, D. and Choi, S. and Lukin, M. D. and Serbyn, M.},
  journal = {Phys. Rev. Lett.},
  volume = {127},
  issue = {9},
  pages = {090602},
  numpages = {7},
  year = {2021},
  month = {Aug},
  publisher = {American Physical Society},
  doi = {10.1103/PhysRevLett.127.090602},
  url = {https://link.aps.org/doi/10.1103/PhysRevLett.127.090602}
}

@article{Hudomal2022,
  title = {Driving quantum many-body scars in the PXP model},
  author = {Hudomal, Ana and Desaules, Jean-Yves and Mukherjee, Bhaskar and Su, Guo-Xian and Halimeh, Jad C. and Papi\ifmmode \acute{c}\else \'{c}\fi{}, Zlatko},
  journal = {Phys. Rev. B},
  volume = {106},
  issue = {10},
  pages = {104302},
  numpages = {19},
  year = {2022},
  month = {Sep},
  publisher = {American Physical Society},
  doi = {10.1103/PhysRevB.106.104302},
  url = {https://link.aps.org/doi/10.1103/PhysRevB.106.104302}
}

@article{Rechtsman2013,
author={Rechtsman, Mikael C.
and Zeuner, Julia M.
and Plotnik, Yonatan
and Lumer, Yaakov
and Podolsky, Daniel
and Dreisow, Felix
and Nolte, Stefan
and Segev, Mordechai
and Szameit, Alexander},
title={Photonic Floquet topological insulators},
journal={Nature},
year={2013},
month={Apr},
day={01},
volume={496},
number={7444},
pages={196-200},
issn={1476-4687},
doi={10.1038/nature12066},
url={https://doi.org/10.1038/nature12066}
}

@article{Aidelsburger2013,
  title = {Realization of the Hofstadter Hamiltonian with Ultracold Atoms in Optical Lattices},
  author = {Aidelsburger, M. and Atala, M. and Lohse, M. and Barreiro, J. T. and Paredes, B. and Bloch, I.},
  journal = {Phys. Rev. Lett.},
  volume = {111},
  issue = {18},
  pages = {185301},
  numpages = {5},
  year = {2013},
  month = {Oct},
  publisher = {American Physical Society},
  doi = {10.1103/PhysRevLett.111.185301},
  url = {https://link.aps.org/doi/10.1103/PhysRevLett.111.185301}
}

@article{Miyake2013,
  title = {Realizing the Harper Hamiltonian with Laser-Assisted Tunneling in Optical Lattices},
  author = {Miyake, Hirokazu and Siviloglou, Georgios A. and Kennedy, Colin J. and Burton, William Cody and Ketterle, Wolfgang},
  journal = {Phys. Rev. Lett.},
  volume = {111},
  issue = {18},
  pages = {185302},
  numpages = {5},
  year = {2013},
  month = {Oct},
  publisher = {American Physical Society},
  doi = {10.1103/PhysRevLett.111.185302},
  url = {https://link.aps.org/doi/10.1103/PhysRevLett.111.185302}
}

@article{Jotzu2014,
author={Jotzu, Gregor
and Messer, Michael
and Desbuquois, R{\'e}mi
and Lebrat, Martin
and Uehlinger, Thomas
and Greif, Daniel
and Esslinger, Tilman},
title={Experimental realization of the topological Haldane model with ultracold fermions},
journal={Nature},
year={2014},
month={Nov},
day={01},
volume={515},
number={7526},
pages={237-240},
issn={1476-4687},
doi={10.1038/nature13915},
url={https://doi.org/10.1038/nature13915}
}

@article{Zhang2017,
author={Zhang, J.
and Hess, P. W.
and Kyprianidis, A.
and Becker, P.
and Lee, A.
and Smith, J.
and Pagano, G.
and Potirniche, I.-D.
and Potter, A. C.
and Vishwanath, A.
and Yao, N. Y.
and Monroe, C.},
title={Observation of a discrete time crystal},
journal={Nature},
year={2017},
month={Mar},
day={01},
volume={543},
number={7644},
pages={217-220},
issn={1476-4687},
doi={10.1038/nature21413},
url={https://doi.org/10.1038/nature21413}
}

@article{Choi2017,
author={Choi, Soonwon
and Choi, Joonhee
and Landig, Renate
and Kucsko, Georg
and Zhou, Hengyun
and Isoya, Junichi
and Jelezko, Fedor
and Onoda, Shinobu
and Sumiya, Hitoshi
and Khemani, Vedika
and von Keyserlingk, Curt
and Yao, Norman Y.
and Demler, Eugene
and Lukin, Mikhail D.},
title={Observation of discrete time-crystalline order in a disordered dipolar many-body system},
journal={Nature},
year={2017},
month={Mar},
day={01},
volume={543},
number={7644},
pages={221-225},
issn={1476-4687},
doi={10.1038/nature21426},
url={https://doi.org/10.1038/nature21426}
}

@article{Rubio-Abadal2020,
  title = {Floquet Prethermalization in a Bose-Hubbard System},
  author = {Rubio-Abadal, Antonio and Ippoliti, Matteo and Hollerith, Simon and Wei, David and Rui, Jun and Sondhi, S. L. and Khemani, Vedika and Gross, Christian and Bloch, Immanuel},
  journal = {Phys. Rev. X},
  volume = {10},
  issue = {2},
  pages = {021044},
  numpages = {14},
  year = {2020},
  month = {May},
  publisher = {American Physical Society},
  doi = {10.1103/PhysRevX.10.021044},
  url = {https://link.aps.org/doi/10.1103/PhysRevX.10.021044}
}

@article{Wintersperger2020,
author={Wintersperger, Karen
and Braun, Christoph
and {\"U}nal, F. Nur
and Eckardt, Andr{\'e}
and Liberto, Marco Di
and Goldman, Nathan
and Bloch, Immanuel
and Aidelsburger, Monika},
title={Realization of an anomalous Floquet topological system with ultracold atoms},
journal={Nature Physics},
year={2020},
month={Oct},
day={01},
volume={16},
number={10},
pages={1058-1063},
issn={1745-2481},
doi={10.1038/s41567-020-0949-y},
url={https://doi.org/10.1038/s41567-020-0949-y}
}

@article{Peng2021,
author={Peng, Pai
and Yin, Chao
and Huang, Xiaoyang
and Ramanathan, Chandrasekhar
and Cappellaro, Paola},
title={Floquet prethermalization in dipolar spin chains},
journal={Nature Physics},
year={2021},
month={Apr},
day={01},
volume={17},
number={4},
pages={444-447},
issn={1745-2481},
doi={10.1038/s41567-020-01120-z},
url={https://doi.org/10.1038/s41567-020-01120-z}
}

@article{Kyprianidis2021,
author={Kyprianidis, A.
and Machado, F.
and Morong, W.
and Becker, P.
and Collins, K. S.
and Else, D. V.
and Feng, L.
and Hess, P. W.
and Nayak, C.
and Pagano, G.
and Yao, N. Y.
and Monroe, C.},
title={Observation of a prethermal discrete time crystal},
journal={Science},
year={2021},
month={Jun},
day={11},
publisher={American Association for the Advancement of Science},
volume={372},
number={6547},
pages={1192-1196},
doi={10.1126/science.abg8102},
url={https://doi.org/10.1126/science.abg8102}
}

@article{Beatrez2023,
author={Beatrez, William
and Fleckenstein, Christoph
and Pillai, Arjun
and de Leon Sanchez, Erica
and Akkiraju, Amala
and Diaz Alcala, Jesus
and Conti, Sophie
and Reshetikhin, Paul
and Druga, Emanuel
and Bukov, Marin
and Ajoy, Ashok},
title={Critical prethermal discrete time crystal created by two-frequency driving},
journal={Nature Physics},
year={2023},
month={Mar},
day={01},
volume={19},
number={3},
pages={407-413},
issn={1745-2481},
doi={10.1038/s41567-022-01891-7},
url={https://doi.org/10.1038/s41567-022-01891-7}
}

@article{Stasiuk2023,
  title = {Observation of a Prethermal $U(1)$ Discrete Time Crystal},
  author = {Stasiuk, Andrew and Cappellaro, Paola},
  journal = {Phys. Rev. X},
  volume = {13},
  issue = {4},
  pages = {041016},
  numpages = {14},
  year = {2023},
  month = {Oct},
  publisher = {American Physical Society},
  doi = {10.1103/PhysRevX.13.041016},
  url = {https://link.aps.org/doi/10.1103/PhysRevX.13.041016}
}

@article{Alessio2014,
  title = {Long-time Behavior of Isolated Periodically Driven Interacting Lattice Systems},
  author = {D'Alessio, Luca and Rigol, Marcos},
  journal = {Phys. Rev. X},
  volume = {4},
  issue = {4},
  pages = {041048},
  numpages = {12},
  year = {2014},
  month = {Dec},
  publisher = {American Physical Society},
  doi = {10.1103/PhysRevX.4.041048},
  url = {https://link.aps.org/doi/10.1103/PhysRevX.4.041048}
}

@article{Lazarides2014,
  title = {Equilibrium states of generic quantum systems subject to periodic driving},
  author = {Lazarides, Achilleas and Das, Arnab and Moessner, Roderich},
  journal = {Phys. Rev. E},
  volume = {90},
  issue = {1},
  pages = {012110},
  numpages = {6},
  year = {2014},
  month = {Jul},
  publisher = {American Physical Society},
  doi = {10.1103/PhysRevE.90.012110},
  url = {https://link.aps.org/doi/10.1103/PhysRevE.90.012110}
}

@article{Ponte2015_1,
title = {Periodically driven ergodic and many-body localized quantum systems},
journal = {Annals of Physics},
volume = {353},
pages = {196-204},
year = {2015},
issn = {0003-4916},
doi = {https://doi.org/10.1016/j.aop.2014.11.008},
url = {https://www.sciencedirect.com/science/article/pii/S0003491614003212},
author = {Pedro Ponte and Anushya Chandran and Z. Papić and Dmitry A. Abanin}
}

@article{Ponte2015_2,
  title = {Many-Body Localization in Periodically Driven Systems},
  author = {Ponte, Pedro and Papi\ifmmode \acute{c}\else \'{c}\fi{}, Z. and Huveneers, Fran\ifmmode \mbox{\c{c}}\else \c{c}\fi{}ois and Abanin, Dmitry A.},
  journal = {Phys. Rev. Lett.},
  volume = {114},
  issue = {14},
  pages = {140401},
  numpages = {5},
  year = {2015},
  month = {Apr},
  publisher = {American Physical Society},
  doi = {10.1103/PhysRevLett.114.140401},
  url = {https://link.aps.org/doi/10.1103/PhysRevLett.114.140401}
}

@article{Lazarides2015,
  title = {Fate of Many-Body Localization Under Periodic Driving},
  author = {Lazarides, Achilleas and Das, Arnab and Moessner, Roderich},
  journal = {Phys. Rev. Lett.},
  volume = {115},
  issue = {3},
  pages = {030402},
  numpages = {5},
  year = {2015},
  month = {Jul},
  publisher = {American Physical Society},
  doi = {10.1103/PhysRevLett.115.030402},
  url = {https://link.aps.org/doi/10.1103/PhysRevLett.115.030402}
}

@article{Abanin2015,
  title = {Exponentially Slow Heating in Periodically Driven Many-Body Systems},
  author = {Abanin, Dmitry A. and De Roeck, Wojciech and Huveneers, Fran\ifmmode \mbox{\c{c}}\else \c{c}\fi{}ois},
  journal = {Phys. Rev. Lett.},
  volume = {115},
  issue = {25},
  pages = {256803},
  numpages = {5},
  year = {2015},
  month = {Dec},
  publisher = {American Physical Society},
  doi = {10.1103/PhysRevLett.115.256803},
  url = {https://link.aps.org/doi/10.1103/PhysRevLett.115.256803}
}

@article{Mori2016,
  title = {Rigorous Bound on Energy Absorption and Generic Relaxation in Periodically Driven Quantum Systems},
  author = {Mori, Takashi and Kuwahara, Tomotaka and Saito, Keiji},
  journal = {Phys. Rev. Lett.},
  volume = {116},
  issue = {12},
  pages = {120401},
  numpages = {5},
  year = {2016},
  month = {Mar},
  publisher = {American Physical Society},
  doi = {10.1103/PhysRevLett.116.120401},
  url = {https://link.aps.org/doi/10.1103/PhysRevLett.116.120401}
}

@article{Kuwahara2016,
title = {Floquet–Magnus theory and generic transient dynamics in periodically driven many-body quantum systems},
journal = {Annals of Physics},
volume = {367},
pages = {96-124},
year = {2016},
issn = {0003-4916},
doi = {https://doi.org/10.1016/j.aop.2016.01.012},
url = {https://www.sciencedirect.com/science/article/pii/S0003491616000142},
author = {Tomotaka Kuwahara and Takashi Mori and Keiji Saito}
}

@article{Abanin2017_1,
  title = {Effective Hamiltonians, prethermalization, and slow energy absorption in periodically driven many-body systems},
  author = {Abanin, Dmitry A. and De Roeck, Wojciech and Ho, Wen Wei and Huveneers, Fran\ifmmode \mbox{\c{c}}\else \c{c}\fi{}ois},
  journal = {Phys. Rev. B},
  volume = {95},
  issue = {1},
  pages = {014112},
  numpages = {8},
  year = {2017},
  month = {Jan},
  publisher = {American Physical Society},
  doi = {10.1103/PhysRevB.95.014112},
  url = {https://link.aps.org/doi/10.1103/PhysRevB.95.014112}
}

@article{Abanin2017_2,
author={Abanin, Dmitry
and De Roeck, Wojciech
and Ho, Wen Wei
and Huveneers, Fran{\c{c}}ois},
title={A Rigorous Theory of Many-Body Prethermalization for Periodically Driven and Closed Quantum Systems},
journal={Communications in Mathematical Physics},
year={2017},
month={Sep},
day={01},
volume={354},
number={3},
pages={809-827},
issn={1432-0916},
doi={10.1007/s00220-017-2930-x},
url={https://doi.org/10.1007/s00220-017-2930-x}
}

@article{Else2017_1,
  title = {Prethermal Phases of Matter Protected by Time-Translation Symmetry},
  author = {Else, Dominic V. and Bauer, Bela and Nayak, Chetan},
  journal = {Phys. Rev. X},
  volume = {7},
  issue = {1},
  pages = {011026},
  numpages = {21},
  year = {2017},
  month = {Mar},
  publisher = {American Physical Society},
  doi = {10.1103/PhysRevX.7.011026},
  url = {https://link.aps.org/doi/10.1103/PhysRevX.7.011026}
}

@article{Mori2018_1,
doi = {10.1088/1361-6455/aabcdf},
url = {https://dx.doi.org/10.1088/1361-6455/aabcdf},
year = {2018},
month = {may},
publisher = {IOP Publishing},
volume = {51},
number = {11},
pages = {112001},
author = {Takashi Mori and Tatsuhiko N Ikeda and Eriko Kaminishi and Masahito Ueda},
title = {Thermalization and prethermalization in isolated quantum systems: a theoretical overview},
journal = {Journal of Physics B: Atomic, Molecular and Optical Physics}
}

@article{Mori2018_2,
  title = {Floquet prethermalization in periodically driven classical spin systems},
  author = {Mori, Takashi},
  journal = {Phys. Rev. B},
  volume = {98},
  issue = {10},
  pages = {104303},
  numpages = {12},
  year = {2018},
  month = {Sep},
  publisher = {American Physical Society},
  doi = {10.1103/PhysRevB.98.104303},
  url = {https://link.aps.org/doi/10.1103/PhysRevB.98.104303}
}

@article{Ho2023,
title = {Quantum and classical Floquet prethermalization},
journal = {Annals of Physics},
volume = {454},
pages = {169297},
year = {2023},
issn = {0003-4916},
doi = {https://doi.org/10.1016/j.aop.2023.169297},
url = {https://www.sciencedirect.com/science/article/pii/S0003491623000829},
author = {Wen Wei Ho and Takashi Mori and Dmitry A. Abanin and Emanuele G. {Dalla Torre}}
}

@article{Bukov2015,
author = {Marin Bukov and Luca D'Alessio and Anatoli Polkovnikov},
title = {Universal high-frequency behavior of periodically driven systems: from dynamical stabilization to Floquet engineering},
journal = {Advances in Physics},
volume = {64},
number = {2},
pages = {139-226},
year  = {2015},
publisher = {Taylor & Francis},
doi = {10.1080/00018732.2015.1055918},
URL = { https://doi.org/10.1080/00018732.2015.1055918},
eprint = { https://doi.org/10.1080/00018732.2015.1055918}
}

@article{Eckardt2015,
doi = {10.1088/1367-2630/17/9/093039},
url = {https://dx.doi.org/10.1088/1367-2630/17/9/093039},
year = {2015},
month = {sep},
publisher = {IOP Publishing},
volume = {17},
number = {9},
pages = {093039},
author = {André Eckardt and Egidijus Anisimovas},
title = {High-frequency approximation for periodically driven quantum systems from a Floquet-space perspective},
journal = {New Journal of Physics}
}

@article{Nathan2021,
title={{Hierarchy of many-body invariants and quantized magnetization in anomalous Floquet insulators}},
author={Frederik Nathan and Dmitry A. Abanin and Netanel H. Lindner and Erez Berg and Mark S. Rudner},
journal={SciPost Phys.},
volume={10},
pages={128},
year={2021},
publisher={SciPost},
doi={10.21468/SciPostPhys.10.6.128},
url={https://scipost.org/10.21468/SciPostPhys.10.6.128},
}

@article{Else2017_2,
  title = {Prethermal Strong Zero Modes and Topological Qubits},
  author = {Else, Dominic V. and Fendley, Paul and Kemp, Jack and Nayak, Chetan},
  journal = {Phys. Rev. X},
  volume = {7},
  issue = {4},
  pages = {041062},
  numpages = {22},
  year = {2017},
  month = {Dec},
  publisher = {American Physical Society},
  doi = {10.1103/PhysRevX.7.041062},
  url = {https://link.aps.org/doi/10.1103/PhysRevX.7.041062}
}

@article{Mikami2016,
  title = {Brillouin-Wigner theory for high-frequency expansion in periodically driven systems: Application to Floquet topological insulators},
  author = {Mikami, Takahiro and Kitamura, Sota and Yasuda, Kenji and Tsuji, Naoto and Oka, Takashi and Aoki, Hideo},
  journal = {Phys. Rev. B},
  volume = {93},
  issue = {14},
  pages = {144307},
  numpages = {25},
  year = {2016},
  month = {Apr},
  publisher = {American Physical Society},
  doi = {10.1103/PhysRevB.93.144307},
  url = {https://link.aps.org/doi/10.1103/PhysRevB.93.144307}
}

@article{Mizuta2019,
  title = {High-frequency expansion for Floquet prethermal phases with emergent symmetries: Application to time crystals and Floquet engineering},
  author = {Mizuta, Kaoru and Takasan, Kazuaki and Kawakami, Norio},
  journal = {Phys. Rev. B},
  volume = {100},
  issue = {2},
  pages = {020301},
  numpages = {5},
  year = {2019},
  month = {Jul},
  publisher = {American Physical Society},
  doi = {10.1103/PhysRevB.100.020301},
  url = {https://link.aps.org/doi/10.1103/PhysRevB.100.020301}
}

@article{Mori2022,
  title = {Heating Rates under Fast Periodic Driving beyond Linear Response},
  author = {Mori, Takashi},
  journal = {Phys. Rev. Lett.},
  volume = {128},
  issue = {5},
  pages = {050604},
  numpages = {6},
  year = {2022},
  month = {Feb},
  publisher = {American Physical Society},
  doi = {10.1103/PhysRevLett.128.050604},
  url = {https://link.aps.org/doi/10.1103/PhysRevLett.128.050604}
}

@article{Morimoto2017,
  title = {Floquet topological phases protected by time glide symmetry},
  author = {Morimoto, Takahiro and Po, Hoi Chun and Vishwanath, Ashvin},
  journal = {Phys. Rev. B},
  volume = {95},
  issue = {19},
  pages = {195155},
  numpages = {16},
  year = {2017},
  month = {May},
  publisher = {American Physical Society},
  doi = {10.1103/PhysRevB.95.195155},
  url = {https://link.aps.org/doi/10.1103/PhysRevB.95.195155}
}

@article{Xu2018,
  title = {Space-Time Crystal and Space-Time Group},
  author = {Xu, Shenglong and Wu, Congjun},
  journal = {Phys. Rev. Lett.},
  volume = {120},
  issue = {9},
  pages = {096401},
  numpages = {6},
  year = {2018},
  month = {Feb},
  publisher = {American Physical Society},
  doi = {10.1103/PhysRevLett.120.096401},
  url = {https://link.aps.org/doi/10.1103/PhysRevLett.120.096401}
}

@article{YP2019,
  title = {Floquet Second-Order Topological Insulators from Nonsymmorphic Space-Time Symmetries},
  author = {Peng, Yang and Refael, Gil},
  journal = {Phys. Rev. Lett.},
  volume = {123},
  issue = {1},
  pages = {016806},
  numpages = {6},
  year = {2019},
  month = {Jul},
  publisher = {American Physical Society},
  doi = {10.1103/PhysRevLett.123.016806},
  url = {https://link.aps.org/doi/10.1103/PhysRevLett.123.016806}
}

@article{YP2020,
  title = {Floquet higher-order topological insulators and superconductors with space-time symmetries},
  author = {Peng, Yang},
  journal = {Phys. Rev. Res.},
  volume = {2},
  issue = {1},
  pages = {013124},
  numpages = {26},
  year = {2020},
  month = {Feb},
  publisher = {American Physical Society},
  doi = {10.1103/PhysRevResearch.2.013124},
  url = {https://link.aps.org/doi/10.1103/PhysRevResearch.2.013124}
}

@article{YP2022,
  title = {Topological Space-Time Crystal},
  author = {Peng, Yang},
  journal = {Phys. Rev. Lett.},
  volume = {128},
  issue = {18},
  pages = {186802},
  numpages = {6},
  year = {2022},
  month = {May},
  publisher = {American Physical Society},
  doi = {10.1103/PhysRevLett.128.186802},
  url = {https://link.aps.org/doi/10.1103/PhysRevLett.128.186802}
}

@article{Na2023,
  title = {Floquet gap-dependent topological classifications from color-decorated frequency lattices with space-time symmetries},
  author = {Na, Ilyoun and Kemp, Jack and Griffin, Sin\'ead M. and Slager, Robert-Jan and Peng, Yang},
  journal = {Phys. Rev. B},
  volume = {108},
  issue = {18},
  pages = {L180302},
  numpages = {6},
  year = {2023},
  month = {Nov},
  publisher = {American Physical Society},
  doi = {10.1103/PhysRevB.108.L180302},
  url = {https://link.aps.org/doi/10.1103/PhysRevB.108.L180302}
}

@misc{SM,
  Note                     = {See Supplementary Material at \url{http://link.aps.org/supplemental/10.1103/3hrs-sd3s} for technical details on Floquet prethermalization in the interaction picture, the van Vleck expansion of the dressed prethermal Hamiltonian, and a rigorous derivation of its symmetry group, along with illustrations of the effective prethermal Hamiltonian and dual-drive Floquet thermalization dynamic, which includes Refs.~\cite{White2018,Rakovszky2022}.}
}

@article{White2018,
  title = {Quantum dynamics of thermalizing systems},
  author = {White, Christopher David and Zaletel, Michael and Mong, Roger S. K. and Refael, Gil},
  journal = {Phys. Rev. B},
  volume = {97},
  issue = {3},
  pages = {035127},
  numpages = {14},
  year = {2018},
  month = {Jan},
  publisher = {American Physical Society},
  doi = {10.1103/PhysRevB.97.035127},
  url = {https://link.aps.org/doi/10.1103/PhysRevB.97.035127}
}

@article{Rakovszky2022,
  title = {Dissipation-assisted operator evolution method for capturing hydrodynamic transport},
  author = {Rakovszky, Tibor and von Keyserlingk, C. W. and Pollmann, Frank},
  journal = {Phys. Rev. B},
  volume = {105},
  issue = {7},
  pages = {075131},
  numpages = {10},
  year = {2022},
  month = {Feb},
  publisher = {American Physical Society},
  doi = {10.1103/PhysRevB.105.075131},
  url = {https://link.aps.org/doi/10.1103/PhysRevB.105.075131}
}

@article{Sameti2019,
  title = {Floquet engineering in superconducting circuits: From arbitrary spin-spin interactions to the Kitaev honeycomb model},
  author = {Sameti, Mahdi and Hartmann, Michael J.},
  journal = {Phys. Rev. A},
  volume = {99},
  issue = {1},
  pages = {012333},
  numpages = {19},
  year = {2019},
  month = {Jan},
  publisher = {American Physical Society},
  doi = {10.1103/PhysRevA.99.012333},
  url = {https://link.aps.org/doi/10.1103/PhysRevA.99.012333}
}

@article{Atala2014,
author={Atala, Marcos
and Aidelsburger, Monika
and Lohse, Michael
and Barreiro, Julio T.
and Paredes, Bel{\'e}n
and Bloch, Immanuel},
title={Observation of chiral currents with ultracold atoms in bosonic ladders},
journal={Nature Physics},
year={2014},
month={Aug},
day={01},
volume={10},
number={8},
pages={588-593},
issn={1745-2481},
doi={10.1038/nphys2998},
url={https://doi.org/10.1038/nphys2998}
}

@article{Sompet2022,
author={Sompet, Pimonpan
and Hirthe, Sarah
and Bourgund, Dominik
and Chalopin, Thomas
and Bibo, Julian
and Koepsell, Joannis
and Bojovi{\'{c}}, Petar
and Verresen, Ruben
and Pollmann, Frank
and Salomon, Guillaume
and Gross, Christian
and Hilker, Timon A.
and Bloch, Immanuel},
title={Realizing the symmetry-protected Haldane phase in Fermi--Hubbard ladders},
journal={Nature},
year={2022},
month={Jun},
day={01},
volume={606},
number={7914},
pages={484-488},
issn={1476-4687},
doi={10.1038/s41586-022-04688-z},
url={https://doi.org/10.1038/s41586-022-04688-z}
}

@article{Machado2019,
  title = {Exponentially slow heating in short and long-range interacting Floquet systems},
  author = {Machado, Francisco and Kahanamoku-Meyer, Gregory D. and Else, Dominic V. and Nayak, Chetan and Yao, Norman Y.},
  journal = {Phys. Rev. Res.},
  volume = {1},
  issue = {3},
  pages = {033202},
  numpages = {13},
  year = {2019},
  month = {Dec},
  publisher = {American Physical Society},
  doi = {10.1103/PhysRevResearch.1.033202},
  url = {https://link.aps.org/doi/10.1103/PhysRevResearch.1.033202}
}

@article{Page1993,
  title = {Average entropy of a subsystem},
  author = {Page, Don N.},
  journal = {Phys. Rev. Lett.},
  volume = {71},
  issue = {9},
  pages = {1291--1294},
  numpages = {0},
  year = {1993},
  month = {Aug},
  publisher = {American Physical Society},
  doi = {10.1103/PhysRevLett.71.1291},
  url = {https://link.aps.org/doi/10.1103/PhysRevLett.71.1291}
}

@article{Else2020_2,
  title = {Long-Lived Interacting Phases of Matter Protected by Multiple Time-Translation Symmetries in Quasiperiodically Driven Systems},
  author = {Else, Dominic V. and Ho, Wen Wei and Dumitrescu, Philipp T.},
  journal = {Phys. Rev. X},
  volume = {10},
  issue = {2},
  pages = {021032},
  numpages = {40},
  year = {2020},
  month = {May},
  publisher = {American Physical Society},
  doi = {10.1103/PhysRevX.10.021032},
  url = {https://link.aps.org/doi/10.1103/PhysRevX.10.021032}
}

@article{He2023,
  title = {Quasi-Floquet Prethermalization in a Disordered Dipolar Spin Ensemble in Diamond},
  author = {He, Guanghui and Ye, Bingtian and Gong, Ruotian and Liu, Zhongyuan and Murch, Kater W. and Yao, Norman Y. and Zu, Chong},
  journal = {Phys. Rev. Lett.},
  volume = {131},
  issue = {13},
  pages = {130401},
  numpages = {8},
  year = {2023},
  month = {Sep},
  publisher = {American Physical Society},
  doi = {10.1103/PhysRevLett.131.130401},
  url = {https://link.aps.org/doi/10.1103/PhysRevLett.131.130401}
}

@article{Gallone2024,
author={Gallone, Matteo
and Langella, Beatrice},
title={Prethermalization and Conservation Laws in Quasi-Periodically Driven Quantum Systems},
journal={Journal of Statistical Physics},
year={2024},
month={Aug},
day={10},
volume={191},
number={8},
pages={100},
issn={1572-9613},
doi={10.1007/s10955-024-03313-9},
url={https://doi.org/10.1007/s10955-024-03313-9}
}

@software{GregDMeyer2024,
  author       = {Gregory D. Kahanamoku-Meyer and
                  Julia Wei},
  title        = {GregDMeyer/dynamite: v0.4.0},
  month        = apr,
  year         = 2024,
  publisher    = {Zenodo},
  version      = {v0.4.0},
  doi          = {10.5281/zenodo.10906046},
  url          = {https://doi.org/10.5281/zenodo.10906046}
}

\end{document}